\documentclass[aps,prr,reprint,a4paper,superscriptaddress,amsmath,amsfonts,amssymb, longbibliography]{revtex4-1}

\usepackage{bm}
\usepackage[caption=false]{subfig}
\usepackage{tikz}
\usetikzlibrary{shapes.misc,positioning,shapes}
\usetikzlibrary{patterns}
\usepackage{array, multirow}
\usepackage[colorlinks,linkcolor=blue,citecolor=blue]{hyperref}%
\usepackage{relsize}

\usepackage{times}

\usepackage[english]{babel}
\usepackage{blindtext}
\newcolumntype{P}[1]{>{\centering\arraybackslash}p{#1}}
\begin{document}
\title{Measurement based estimator scheme for continuous quantum error correction}
% \title{Always-on quantum error correction using a measurement driven stochastic feedback controller} # There are already a few articles by this name
\date{\today}
% \date{Phys. Rev. Research 4, 033207, 2022}

\author{Sangkha Borah}
\email{sangkha.borah@oist.jp}
\affiliation{Okinawa Institute of Science and Technology Graduate University, Onna-son, Okinawa 904-0495, Japan}
\author{Bijita Sarma}
% \email{bijita.sarma@oist.jp}
\affiliation{Okinawa Institute of Science and Technology Graduate University, Onna-son, Okinawa 904-0495, Japan}
\author{Michael Kewming}
% \email{m.kewming@uq.edu.au}
\affiliation{Department of Physics, Trinity College Dublin, Dublin 2, Ireland}
\author{Fernando Quijandr\'{\i}a} 
\affiliation{Okinawa Institute of Science and Technology Graduate University, Onna-son, Okinawa 904-0495, Japan}
\author{Gerard J. Milburn}
% \email{g.milburn@uq.edu.au}
\affiliation{Centre for Engineered Quantum Systems, School of Mathematics and Physics, University of Queensland, QLD 4072 Australia}
\author{Jason Twamley}
% \email{jason.twamley1@oist.jp}
\affiliation{Okinawa Institute of Science and Technology Graduate University, Onna-son, Okinawa 904-0495, Japan}

%TC:ignore
\begin{abstract}
Canonical discrete quantum error correction (DQEC) schemes use projective von Neumann measurements on stabilizers to discretize the error syndromes into a finite set, and fast unitary gates are applied to recover the corrupted information. QEC based on continuous measurement (CQEC), in principle, can be executed faster than DQEC and can also be resource efficient. However, CQEC requires meticulous filtering of noisy continuous measurement data to reliably extract error syndromes on the basis of which errors could be detected.  In this work, we show that by constructing a measurement-based estimator (MBE) of the logical qubit to be protected, which is driven by the noisy continuous measurement currents of the stabilizers, it is possible to accurately track the errors occurring on the physical qubits in real-time.  We use this MBE to develop a novel continuous quantum error correction (MBE-CQEC) scheme that can protect the logical qubit to a high degree, surpassing the performances of DQEC, and also allows QEC to be conducted either immediately or in delayed time with  instantaneous feedbacks.
\end{abstract}
\maketitle
%TC:endignore
Generally speaking, quantum error correction (QEC) is a  solution to preserve a quantum state from environmental decoherence, and is essential for achieving fault-tolerant quantum computation, cryptography and quantum communications~\cite{NielsenChuangBook, Shor1995Oct, Steane1996Jul, Djordjevic2012, Gertler2021Feb}. The essence of QEC is to redundantly encode the quantum information of a qubit in several entangled qubits which collectively form a so called logical qubit that exhibits a longer lifetime than individual component physical qubits. The logical qubit lies in a two-dimensional subspace of the Hilbert space of the physical qubits, and the interaction between the qubits and their environment cause an orthogonal rotation of the collective state of the physical qubits out of this subspace. By simultaneously measuring a set of operators, this rotation can be detected and corrected without changing the encoded logical qubit state.  Such operators are selected parity operators in the Pauli group, called the stabilizer generators,  the eigenvalues of which are known as the error syndromes~\cite{Shor1995Oct, Steane1996Jul, Gottesman1997May}. In canonical QEC methods, which we will refer to as discrete quantum error correction (DQEC), these operators are measured projectively and reveal the discrete error syndromes and this classical information is subsequently used to correct qubit errors via fast unitary gates~\cite{Devitt2013Jun, Reed2012Feb}. 
To achieve fault tolerant quantum computation, it is important that the probability of an erroneous rotation of the logical qubit, is below a  critical threshold value~\cite{Knill1998Jan, Girvin2021Nov}.  Over the few years, DQEC has been demonstrated experimentally in various platforms such as in ion traps~\cite{ Schindler2011May, Linke2017Oct,  Negnevitsky2018Nov}, diamond NV centers~\cite{Cramer2016May}, and superconducting circuits~\cite{Kelly2015Mar, Riste2015Apr, Ofek2016Aug, Arute2019Oct, Andersen2020Aug, Stricker2020Sep, Chen2021Jul, deNeeve2022Mar}.

A less explored alternative to DQEC  is to utilize continuous quantum error correction (CQEC) methods, with the first few studies dating back to the early 2000s, co-authored by one of the authors of this article~\cite{Ahn2002Mar,Ahn2003May, Sarovar2004May, Sarovar2005Jul, WisemanMilburnBook}. 
In CQEC, instead of discrete projective measurements of the stabilizer generators, these generators are continuously and weakly measured and a quantum feedback control Hamiltonian is used for continuous error correction.  One early seminal result  demonstrated how single bit flip errors can be corrected using CQEC provided  one knows the conditional moments of the error syndromes, which alas, is not  practically feasible ~\cite{Ahn2002Mar}. This is because when we perform a weak measurement on the stabilizers, we no longer have the direct access to the exact syndrome signals, since they are now masked by the measurement noise that is necessarily added to the measured signal. Likewise, in previous research along these directions, the continuous measurement records of the syndrome measurements were smoothed with various filter kernels so that the exact signal of the error syndromes could be extracted from the noisy measurement records. Expectedly, this performed suboptimally given  that it is not  possible to isolate the signal from  noise for any realistic situation  ~\cite{Sarovar2004May, Mabuchi2009Oct, Cardona2019Jan, Mohseninia2020Nov,  Atalaya2021Apr, Livingston2021Jul}. Following a similar strategy for filtering noisy data, CQEC has been demonstrated experimentally for the first time in a superconducting circuit platform last year~\cite{Livingston2021Jul}.  

Unlike DQEC, which relies on projective measurements, CQEC eliminates the need to use ancilla qubits to measure the stabilizer operators by weakly measuring the physical qubits, and allows faster measurements and error detection, thereby greatly reducing the likelihood of undetected errors~\cite{WisemanMilburnBook}. Furthermore, CQEC can be advantageous when the control resources are limited and the performance of the feedback can be improved by optimizing the operational parameters~\cite{Sarovar2004May}. However, as mentioned above, previous methods to perform CQEC suffered from their inability to correctly identify when errors occur as the continuous measurement signals necessarily contain noise~\cite{Sarovar2004May, Atalaya2021Apr, Livingston2021Jul}. In this work we show how to overcome this and push the capabilities of the CQEC far beyond the abilities of the standard DQEC. To achieve this we equip the CQEC with a real-time  \textit{Measurement Based Estimator (MBE)}, which can detect and correct errors rapidly without filtering or smoothing of the measured data. We call this scheme of continuous error correction as \textbf{MBE-CQEC}. This gives a practical solution for realizing the theoretical proposal of Ahn et al.~\cite{Ahn2002Mar}, without using filters for signal processing of the measurement records~\cite{Sarovar2004May, Livingston2021Jul}. Finally, we show that the corrective action  need not be instantaneous, but can be delayed and corrected whenever required, a feature we call \textit{delayed error correction (DEC)}.

% \section{Results}
\begin{figure}[t]
    \centering
    \includegraphics[width=1.0\linewidth]{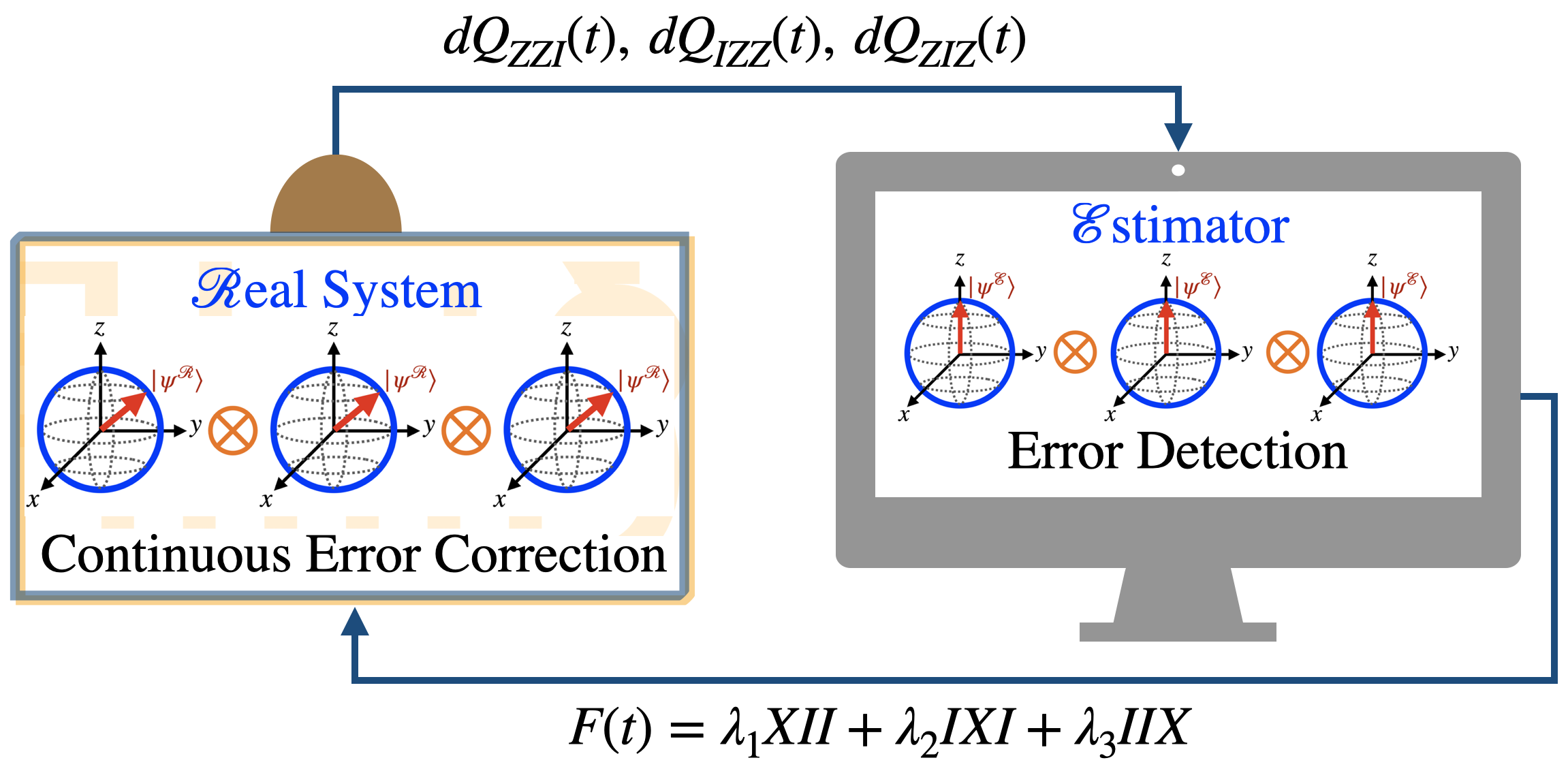}
    \caption{The proposed protocol for CQEC using the measurement based $\cal E$stimator (MBE) scheme. The $\cal R$eal system (left) consists of a logical qubit comprising three physical qubits with an encoded unknown quantum state $|\psi^{\cal R}\rangle_L = \alpha |000\rangle + \beta |111\rangle$, which we want to protect from bit-flip errors. The $\cal E$stimator (right) is a simulation [computer] of the stochastic dynamics of the $\cal R$eal system modelled similarly but with a different initial quantum state $|\psi^{\cal E}\rangle_L = \alpha^\prime |000\rangle + \beta^\prime |111\rangle$, where $\alpha^\prime \neq \alpha$ and $\beta^\prime \neq \beta$ and where $[\alpha,\beta](\alpha^\prime, \beta^\prime)$ are [unknown]known. For generality, we will initialize the $\cal E$stimator state at: $|\psi^{\cal E}\rangle_L = 1|000\rangle$. One executes separate continuous measurements of the three syndrome generators on the $\cal R$eal system and the resulting  time varying classical signals $(dQ_{ZZI}(t),dQ_{IZZ}(t),dQ_{ZIZ}(t))$  drive the stochastic dynamics of the $\cal E$stimator quantum dynamics. Although the $\cal E$stimator cannot learn about the unknown encoded quantum state, any errors appearing in the $\cal R$eal system are faithfully reproduced in the $\cal E$stimator. By monitoring the appearance of bit-flips in the $\cal E$stimator one applies a feedback Hamiltonian $F(t)$ which applies the appropriate correction in a continuous manner with control strengths $\lambda_j$ on {the} individual physical qubits {in the $\cal R$eal system}. 
    }
    \label{fig:fig1}
\end{figure}
\begin{figure*}[!hbt]
    \centering
    \includegraphics[trim={1.60cm 0.35cm0 2.2cm0 0.75cm0},clip, width=1\linewidth]{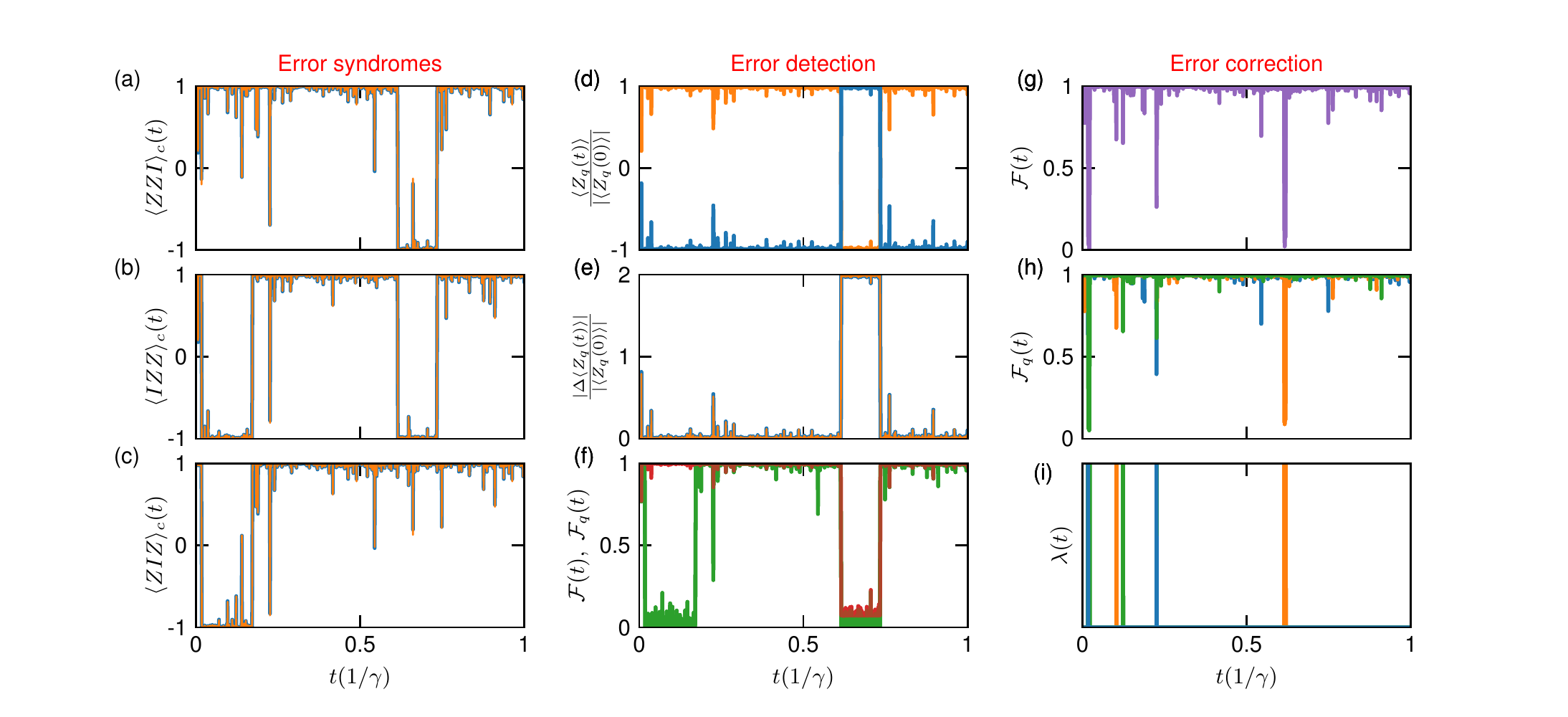}
    \caption{[(a) - (c)] The time evolution of the conditional means of the three error syndromes for the $\cal R$eal system (in blue, linewidth slightly increased for visibility) and the $\cal E$stimator (orange) under the MBE scheme (without error correction) that explains the essence of the protocol of the proposed CQEC scheme. {Essentially, the MBE scheme allows us to use a computer simulation driven by the continuous measurement currents of the $\cal R$eal system to perform real-time quantum error tracking that permits real time QEC} (d) {Instead of using the conditional means of the syndromes we find that the MBE has access to the full real-time effects of errors on each simulated physical qubit and this information can permit us to perform error correction.} Information about the evolution of  the Pauli-$Z_q$ operator for the physical qubit $q$, $\langle Z_q(t) \rangle$, scaled by its initial absolute value $|\langle Z_q(0) \rangle|$ for one of the physical qubits of the $\cal E$stimator (orange) is compared, as an example, to the corresponding evolution of the same of the $\cal R$eal system (blue). (e) The absolute values of the
    instantaneous differences $\Delta \langle Z_q(t) \rangle = \langle Z_q(t) \rangle - \langle Z_q(0) \rangle$ scaled by their initial values $\langle Z_q(0) \rangle$ follow one another for the respective qubits of the $\cal R$eal and the $\cal E$stimator. (f) The fidelity of the particular physical qubit, ${\cal F}_q$ (red) as well as of the logical qubit, ${\cal F}$ (green) with respective to the initial state to be preserved is shown for the bit-flip errors demonstrated in [(a) - (e)]. Thus, the change in fidelity of the physical qubits can be directly monitored by computing the values of $|\langle Z_q(t) \rangle - \langle Z_q(0) \rangle|$ for each of the qubits of the $\cal E$stimator, which sets the bit-flip error detection protocol of the proposed MBE-CQEC scheme. (g) The performance of the MBE-CQEC scheme is showcased in terms of the logical qubit fidelity for a single trajectory to show how the errors are corrected once they are detected, based on the above error detection protocol; also showing the fidelities of each physical qubits, in (i), and the feedbacks on the individual qubits to correct the corrupted information at appropriate times, in (j). For these analyses, we set the initial logical qubit state as $|\psi^{\cal R}\rangle_L = |111\rangle$ to maximize the contrast of fidelity drop under bit-flip error. For these plots, we have used $\kappa/\gamma = 800$ and $\lambda/\gamma = 600$. 
    }
    \label{fig:fig2}
\end{figure*}

The generalized MBE is constructed as follows. Let us consider that the internal dynamics of this $\cal R$eal system is described by the Hamiltonian $H$, and its conditional density matrix to be, $\rho^{\cal R}_c(t)$, under continuous measurement of the operator $A^\mathcal{R} = A$. This can be described by the quantum stochastic master equation (SME)~\cite{WisemanMilburnBook, Diosi2006Sep, Borah2021Nov}, 
\begin{align}
\nonumber
d\rho^{\cal R}_c(t) = & - i [H, \rho^\mathcal{R}_c(t)] dt +  \gamma \mathcal{D}[c] \rho^{\cal R}_c(t) dt + \kappa \mathcal{D}[A] \rho^{\cal R}_c(t) dt \\
+& \sqrt{\kappa\eta} \mathcal{H}[A] \rho^{\cal R}_c(t)\, dW^{\cal R}(t).
\end{align}
Here, $A = \mathcal{A}/\mathcal{A}_0$ a dimensionless operator corresponding to the physical observable $\mathcal{A}$ scaled suitably by $\mathcal{A}_0$ to make it dimensionless, and is known as the measurement operator, which is measured with a measurement rate of $\kappa$. The first term on the RHS is the coherent evolution of the system. The second term on the RHS gives the environmental decoherence at a rate $\gamma$ with the collapse operator $c$, and the third term  gives the measurement backaction due to the measurement of $A$, where $\mathcal{D}[A]\rho = A\rho A^\dagger - \frac{1}{2}(A^\dagger A \rho + \rho A^\dagger A)$ represents the decoherence superoperator. The last term is the stochastic diffusion term with $dW(t)$ being the Wiener noise increments. $\mathcal H$ is a superoperator given by, $\mathcal{H}[A]\rho = A\rho + \rho A^\dagger - \rho \mathrm{tr}[A\rho + \rho A^\dagger]$, and $\eta \in (0, 1]$ is the measurement efficiency. The measurement records, $dQ^\mathcal{R}(t)$ are given by the summation of the conditional mean of the measurement operator and the corresponding random noise component of the measurement, 
\begin{align}
dQ^\mathcal{R}(t) = \langle A^\mathcal{R} (t) \rangle_c dt + \frac{1}{\sqrt{4\kappa\eta}} {dW^\mathcal{R}(t)}. 
\end{align}
The dynamics of the $\cal E$stimator is modelled following the modified SME,
\begin{align}
\nonumber
d\rho^{\cal E}_c(t) = & - i [H, \rho^{\cal E}_c(t)] dt +  \gamma \mathcal{D}[c] \rho^{\cal E}_c(t) dt 
+ \kappa \mathcal{D}[A] \rho^{\cal E}_c(t) dt \\
+& {2\kappa\eta}\left[
dQ^{\cal R}(t) - \langle A^{\cal E} (t) \rangle_c dt \right] \mathcal{H}[A] \rho^{\cal E}_c(t) .
\label{eq:mbe_scheme_main_eq}
\end{align}
In essence, the noise of the $\cal E$stimator is modelled based on the noisy measurement records of the $\cal R$eal system. In the context of the present work for correcting bit-flip errors of the three-qubit code, we would have, $c = \{XII, IXI, XIX \}$ and $A = \{ZZI, IZZ, ZIZ \}$.

The proposed MBE-CQEC scheme is shown schematically in Fig.~\ref{fig:fig1}. The $\cal R$eal system (left) with the initial state $\rho^{\cal R}$ consists of a logical qubit comprising three physical qubits we take to be a system in the laboratory, where one continuously measures the stabilizer operators $ZZI, IZZ$ and $ZIZ$, where the third stabilizer operator is redundant and can be omitted in principle. We consider an $\cal E$stimator of the system with the initial state $\rho^{\cal E}$, as a numerical simulator on a fast computer (on the right in Fig.~\ref{fig:fig1}), whose purpose is to detect the errors on the qubits occurring in the $\cal R$eal system based on the real time measurement records. The $\cal E$stimator also acts as a controller to apply an appropriate feedback Hamiltonian to the $\cal R$eal system: $F(t) = \lambda_1(t) XII + \lambda_2(t) IXI + \lambda_3(t) IIX$, where $X$ denotes a Pauli-$X$ operator, and $\lambda_q(t)$'s are the feedback strengths. 

At the heart of the MBE-CQEC lies the fact that for the measurement of the stabilizer operators, the $\cal E$stimator can perfectly follow the conditional means of the stabilizers ($\langle ZZI \rangle_c(t), \langle IZZ \rangle_c(t)$ and $\langle ZIZ \rangle_c(t)$) when it is fed the continuous, albeit noisy syndrome measurement records ($dQ_{ZZI}(t),dQ_{IZZ}(t),dQ_{ZIZ}(t)$). In Fig.~\ref{fig:fig2}[(a) - (c)], we show these for the $\cal R$eal (blue, with slightly thicker lines for visibility) and the $\cal E$stimator (orange) for a single quantum trajectory, where the $\cal E$stimator dynamics is driven by the syndrome measurement currents of the $\cal R$eal system.  The perfect match of these values entails the power of the approach  we are going to formulate for QEC, which offers a novel strategy to extract the {conditional means of the } error syndromes using continuous measurement instead of projective von Neumann measurement {and without any signal filtering}.  This would allow CQEC to operate at the optimum level of performance using the error syndromes directly while outperforming the DQEC protocols~\cite{Ahn2002Mar, Sarovar2004May, WisemanMilburnBook, NielsenChuangBook}, thus making it a perfect marriage between DQEC and CQEC techniques.

\begin{figure*}[!hbt]
    \centering
    \includegraphics[trim={1cm 0.75cm0 1cm0 1.0cm0},clip, width=1.0\linewidth]{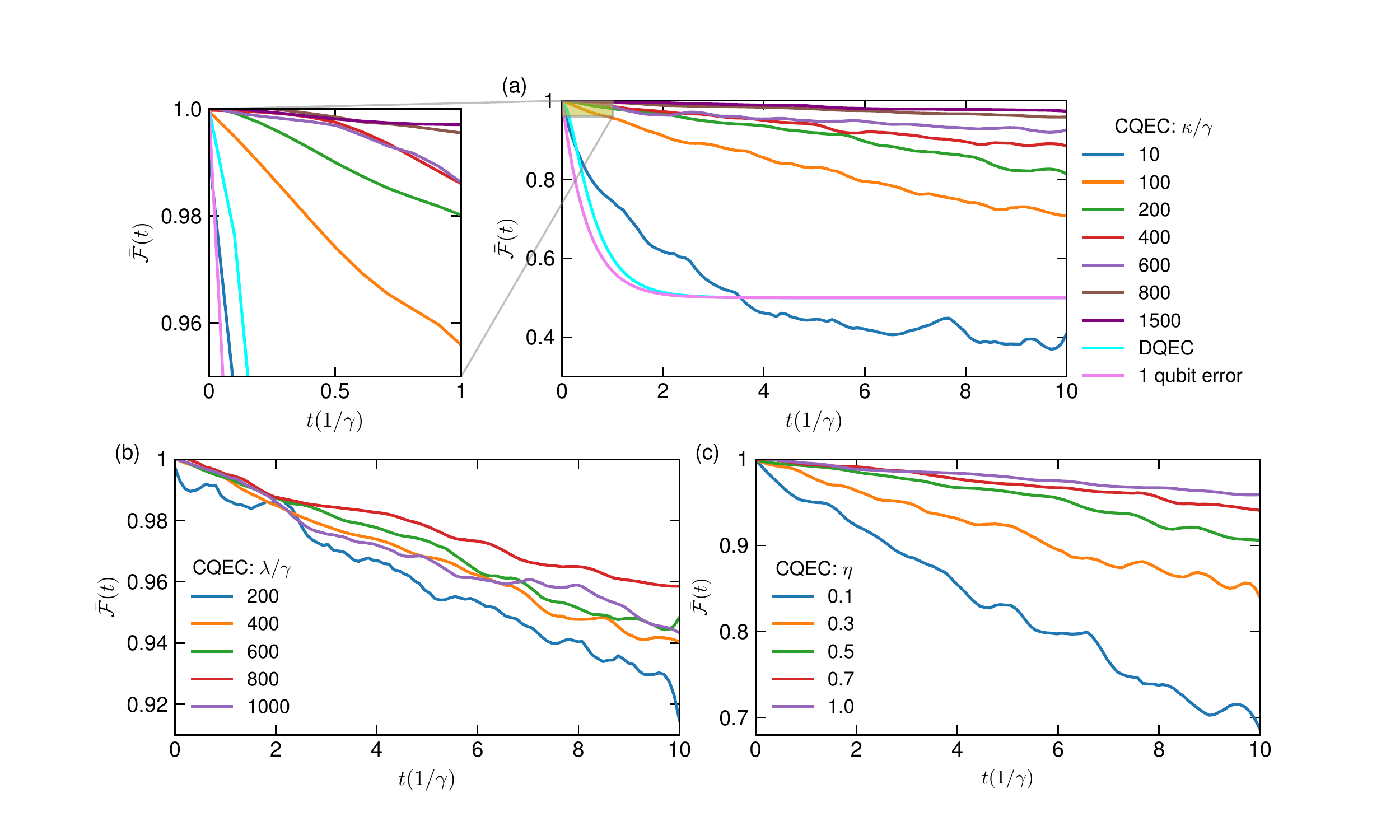}
    \caption{(a) The net fidelity of the logical qubit averaged over an ensemble of at least 100 trajectories  under the proposed MBE-CQEC scheme for different choices of measurement strength, $\kappa$ in the units of the qubit bit flip error rate $\gamma$ are shown with the feedback strengths $\lambda = \kappa$, for reasons demonstrated in (b). The time evolutions up to 10 lifetimes of the qubit ($10/\gamma$) are corrected. Also shown are the {fidelities for the } DQEC (cyan) and one (violet) physical qubit errors  for comparison.  On the left of (a) the zoomed-in portion within $t = 1/\gamma$ is shown. While DQEC fails completely beyond a few lifetimes of the qubit (one qubit error), MBE-CQEC protocol outperforms it {significantly}. (b) The performance of CQEC scheme for different choices of the feedback strength, $\lambda$ for a fixed $\kappa/\gamma = 800$. It shows that $\lambda \sim \kappa$ is a decent choice for overall high fidelity in long time limit. (c) The performance of the scheme for non-ideal choices of measurement efficiencies, $\eta$, showing that the drop in fidelity relative to the case of ideal measurement efficiency, $\eta = 1$ is not significantly large {for reasonable values of $\eta$.}}
    \label{fig:fig3}
\end{figure*}

\begin{figure}[!hbt]
    \centering
    \includegraphics[trim={0.5cm 0.05cm0 1.0cm0 0.5cm0},clip, width=1.0\linewidth]{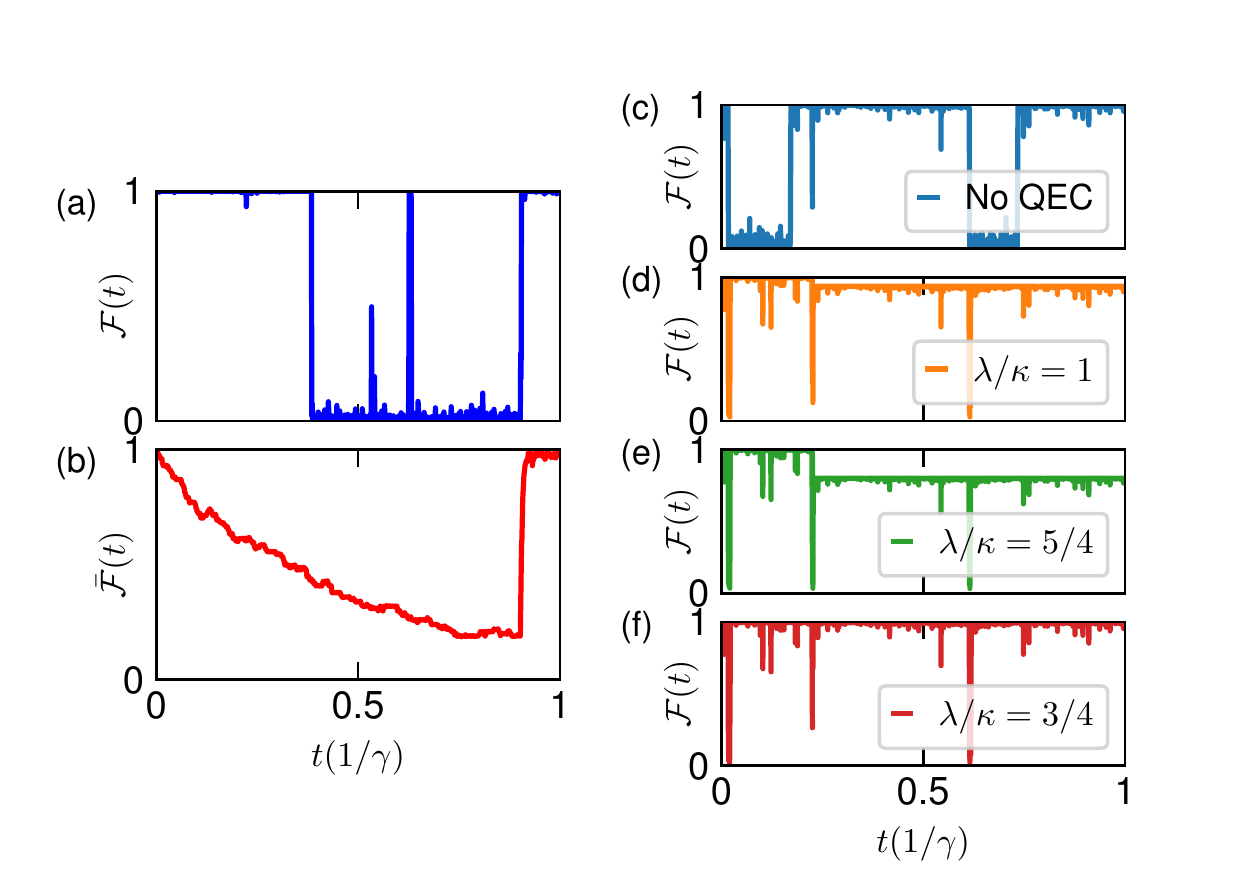}
    \caption{(a) and (b) demonstrate the delayed error correction (DEC) allowed by the MBE-CQEC scheme, where the error correction can be deferred until a later time.  In these simulations, we do not apply any error correction till $t=0.9/\gamma$, after which the errors are detected based on the proposed error detection scheme and faithfully corrected. (a) An example trajectory shows how the error is corrected at  $t=0.9/\gamma$.  (b) The same as (a) but averaged over many trajectories, which shows how the overall fidelity drops until the errors are corrected. [(c) - (f)] Explanation of the fidelity drop with the CQEC protocol based on the MBE scheme: (c) the fidelity variation in time without CQEC ($\lambda = 0$) is shown for a particular quantum trajectory for $\kappa/\gamma = 800$; (d) the same when used for CQEC using $\lambda = \kappa$, that reveals that the error at $t \sim 0.2/\gamma$ could not be corrected fully; (e) the same as (d) but for $\lambda = 5\kappa/4$, for which the error correction is even worse at that particular instance; (f) the errors getting perfectly corrected for $\lambda = 3\kappa/4$. 
    }
    \label{fig:fig4}
\end{figure}

While the perfect computation of the {real time conditional} error syndromes with the proposed MBE scheme provides optimal error correction with continuous measurements, we will show in the following that we can use yet another error detection protocol that depends on the time evolution of the Pauli-$Z$ operator of the $\cal E$stimator physical qubits $q$, conditioned on the measurement records of the $\cal R$eal system. This protocol will have the benefit of performing CQEC with a significant time delay solely based on measurement data obtained from the real qubit measurements, and will have additional benefits, which we will discuss in subsequent sections. In Fig.~\ref{fig:fig2}(d), we compare, the time evolution of the $\langle Z_q (t) \rangle$  (only one of the physical qubits is shown as an example) of the instantaneous density matrix of the $\cal R$eal, $\rho^{\cal R}(t)$ and $\cal E$stimator, $\rho^{\cal E}(t)$ where the $\cal E$stimator is evolved according to the MBE scheme discussed above. As expected, the expectation value $\langle Z_q(t) \rangle$ for the $\cal R$eal system (in blue) and the $\cal E$stimator (in orange) exhibit different values, as the initial states are different. However, when normalized by their absolute values before measurement, $\langle Z_q(0) \rangle$, they undergo similar changes. As shown in Fig.~\ref{fig:fig2}(e), the instantaneous differences $\Delta \langle Z_q(t) \rangle = \langle Z_q(t) \rangle - \langle Z_q(0) \rangle$ scaled by their initial values $\langle Z_q(0) \rangle$ follow one another for the respective qubits of the $\cal R$eal and the $\cal E$stimator. In Fig.~\ref{fig:fig2}(f), the fidelity of that particular physical qubit $q$, ${\cal F}_q(t)=\langle \psi(0), \mathrm{tr}_q(\rho(t)) \psi(0)\rangle$ (red), where $\mathrm{tr}_q(\rho(t))$ is the partial trace of the logical qubit density matrix on the $q$\textsuperscript{th} Hilbert space, along with the codespace fidelity of the logical qubit, ${\cal F}(t)=\langle \psi^L(0), \rho(t) \psi^L(0)\rangle$ (green) are shown, from which, by comparing with Fig.~\ref{fig:fig2}(e), it is observed that the drop in fidelity of the individual qubits are directly related in a one to one fashion to $|\Delta \langle  Z_q(t)\rangle/|\langle  Z_q(0)\rangle|$ of the $\cal R$eal/$\cal E$stimator systems. This fact can be utilized to  detect flipping of the qubits deterministically, as an error in a qubit would mean simply $|\Delta \langle Z_q(t)\rangle| > \varepsilon |\langle Z_q(0) \rangle|$ on the $\cal E$stimator, where $\varepsilon = 1.05$ is a small tolerance to error detection.  Naturally, $\varepsilon = 2$ would signify a complete flip, while $\varepsilon = 0$ would signify a complete preservation of the state. For the $\cal E$stimator, we can fix the {initial} logical state of the qubit at the reset conditions, $|\psi\rangle_L^{\cal E} = |000\rangle$ for simplicity and generality. For more details of the error detection protocol, see Methods.

Using the above approach of error detection, we next go on to the implementation of the MBE-CQEC protocol.  In Fig.~\ref{fig:fig2}(g), we demonstrate our CQEC scheme by applying it to a quantum trajectory {evolved over} one lifetime of {a physical} qubit ($t = 1/\gamma$). It can be seen how well the scheme works in correcting the bit-flip errors, quickly {restoring the logical qubit} after an error is detected, as detected by the error syndromes in Fig.~\ref{fig:fig2}(a)-(c). In Fig.~\ref{fig:fig2}(h), the individual fidelities of the physical qubits are shown, and the times of the applied feedbacks on respective qubits are shown in (i). There is hardly any drop in fidelity for this particular trajectory under MBE-CQEC within {this} time span.   

In order to evaluate the performance of the scheme correctly in a statistical sense, we apply it to an ensemble of quantum trajectories and average over it, the results of which are shown in Fig.~\ref{fig:fig3} in terms of average codespace fidelity, $\bar{\cal F}=1/N\sum_{j=1}^N \mathcal{F}_i$ of the logical qubit, where $N$ represents the number of trajectories for the ensemble average, and $\mathcal{F}_i$ is the codespace fidelity of the $i$\textsuperscript{th} trajectory, as defined earlier. While the feedback strengths $\lambda$'s can be tuned in principle within a trajectory, we have used a constant value, $\lambda_0 \sim \kappa$ for simplicity, which means that $\lambda$ can only take the values of $0$ or $\lambda_0$.  In Fig.~\ref{fig:fig3}(a), we show how higher values of the {continuous} measurement rate $\kappa$ yields an overall higher fidelity in the long time limit for the encoded state. The time limit considered is 10 times the {single} qubit lifetime ($10 / \gamma$). We have found that {by choosing a} feedback strength of $\lambda \sim \kappa$ is a good choice that leads to an overall higher fidelity when averaged over hundreds of trajectories. The performance of DQEC is also shown for comparison, which shows that the MBE-CQEC scheme outperforms {DQEC} for $\kappa > 10\gamma$. Another useful measure that is typically checked and useful for fault-tolerance is the so-called one qubit error fidelity, shown as a solid line in violet, which saturates at ${1/2}$ in {the} long time limit. DQEC fails completely beyond about $t \sim \pi/\gamma$, whereas with the MBE-CQEC scheme, the {infidelity} is maintained within $1-4\%$  for higher values of $\kappa$ over 10  lifetimes of a physical qubit. The zoomed in view of the plot, focussing on the performance within 1 lifetime of a physical qubit, is shown on the left of Fig.~\ref{fig:fig3}(a). The final fidelities for $\kappa = 1000\gamma$ and $800\gamma$ are respectively $99.857\%$ and $99.578\%$ at $t=1/\gamma$.  Thus, it is not necessarily useful to keep pushing the value of $\kappa$ any further, as the above  fidelities are not significantly different in magnitude. Recent developments of quantum technologies, particularly those based on superconducting hardwares, allow experimentalists to go beyond the conventional regimes of weak coupling~\cite{Minev2019Jun, Blais2021May, Livingston2021Jul}. In this spirit, we will consider $\kappa/\gamma = 800$ for further analyses in the rest of the paper. In Fig.~\ref{fig:fig3}(b), we evaluate the proposed CQEC scheme for different choices of $\lambda/\gamma$ while keeping $\kappa/\gamma = 800$. It reveals that $\lambda \sim \kappa$ is a relatively decent choice to preserve overall fidelities in the long time limit. Next, we evaluate the performance of the protocol for inefficient measurements ($\eta < 1$), shown in Fig.~\ref{fig:fig3}(c) relative to the ideal case, $\eta = 1$. It is observed that the scheme is fairly robust for $\eta > 0.5$, and the drop of fidelity is not huge.   

Now, we will discuss a distinct feature of the proposed MBE-CQEC scheme facilitated by the new error-detection/correction scheme discussed above.  We find that we can delay the correction until some later time when it is more convenient. We call this feature delayed error correction (DEC). This is facilitated by the proposed error detection scheme based on computation of the Pauli-$Z$ operator on the physical qubit $q$ of the $\cal E$stimator model under the MBE scheme. This allows to keep a track the changes of $|\langle Z_q(t) \rangle - \langle Z_q(0) \rangle|/|\langle Z_q(0) \rangle|$ on the $\cal R$eal system qubits indirectly by monitoring the same on the $\cal E$stimator qubits (see discussions at the beginning, and Methods). For instance, within a trajectory of total time of $1/\gamma$, we can abstain from doing any error correction until a later time, say $t = 0.9/\gamma$. Based on the measurement records of the $\cal R$eal system, the $\cal E$stimator can follow the errors that happened on the qubits, and the errors can be rightfully detected and corrected using the proposed MBE-CQEC protocol, shown for an example trajectory in Fig.~\ref{fig:fig4}(a). The same for an ensemble of trajectories is shown in Fig.~\ref{fig:fig4}(b). This shows how the fidelity drops significantly to a very low value without error correction, but how the error is corrected instantly at $t = 0.9/\gamma$ just monitoring the $\cal E$stimator $\langle Z(t)\rangle$ on individual qubits.

Finally, we will address the long time fidelity drop issue despite error correction using our MBE-CQEC code, and possibilities of using fine-tuned feedback controls. We have found that the choice of the feedback strength $\lambda$ plays a key role in maintaining codespace fidelity for longer durations. To demonstrate it, we first simulate a trajectory that undergoes bit-flip errors as shown in Fig.~\ref{fig:fig4}(c) and save the noise signal, in order to test the effect of different feedback strengths on exactly the same dynamics.  For this simulation, we use $\kappa/\gamma = 800$ as before.  Now, we use our CQEC scheme with $\lambda = \kappa$, for which we see that the fidelity could not be preserved beyond $t \sim 0.2/\gamma$, shown in Fig.~\ref{fig:fig4}(d). Applying the same for  $\lambda = 5\kappa/4$ leads to further drop in fidelity (Fig.~\ref{fig:fig4}(e)) at that point. This can be corrected perfectly, however, with $\lambda = 3\kappa/4$ (Fig.~\ref{fig:fig4}(f)). Hence, the reason for the drop in fidelity with time can be attributed to the instantaneous choices of the feedback strengths. Although, in the discussion for Fig.~\ref{fig:fig3}(c), we had found that $\lambda \sim \kappa$ serves as a decent choice of feedback strength, in principle fine tuning it will improve the achievable fidelity in the long time limit. Optimizing the values of $\lambda_j(t)$ however, may not be a trivial task. 

In this work, we have formulated an innovative approach of realizing bit-flip QEC that can be regarded as one of the most optimal error correcting methods in the literature, which is named as measurement based estimation controlled continuous quantum error correction (MBE-CQEC). While traditionally used methods of QEC are based on projective measurements, the current proposal utilizes continuous measurements at its core, and thus falls under CQEC. While CQEC, in principle, can be carried out in much quicker time intervals, the biggest problem is the absence of the true error syndromes as these signals get masked by the measurement noises. In our proposed method of CQEC, based on a measurement based estimation scheme, the best of both QEC and traditional CQEC techniques could be achieved. In addition, a novel bit-flip error detection scheme was formulated that can be operated in delayed time.  In practical scenarios, each gate carries intrinsic error, which albeit being small, accumulates in time and poses a challenge to achieve fault-tolerant quantum computation. The delayed CQEC method can be advantageous in this particular context, where the intervals of successive error correction steps can be kept significantly high. Also note that for the analysis of the work, we assume that the encoding was done perfectly. However, MBE-CQEC is expected to be resilient to small encoding errors thanks to the perfect emulation of the individual qubit errors and the way the errors are detected based on Pauli-$Z$ expectation value deviation.

Finally, while the method works optimally with the distinctive features of delayed QEC, the bottleneck would come from numerical expenses when one would try to extend to more qubits, as the Hilbert space dimension would grow as $2^N$, where $N$ is the number of qubits. In addition, for best performance the detector should exhibit high response bandwidth. While phase-shift errors can be corrected as bit-flip errors by moving to the computational basis of qubits, the inclusion of both errors in a single code, e.g., the 9-qubit Shor code, will be limited drastically by the computational effort required to solve the $\cal E$stimator dynamics in real time. In this context, the use of a compact representation of the states, e.g. the matrix product states, could be useful.

In conclusion, we have proposed a novel approach of doing bit-flip error correction that performs optimally. It not only removes the limitations of canonical projective QEC techniques, but also can be used to correct errors in delayed time based on all the previous measurement records, which can be a much welcoming factor for its experimental realization.

\section*{Appendix}

\textbf{Quantum continuous measurement and feedback control.}
Contrary to von Neumann measurements, where a measurement operator (observable) is projectively measured collapsing the state to an eigenstate, continuous measurements are weak measurements where the observable is monitored in real time without perturbing the state of the system significantly, so that the collapse is gradual. These types of measurements are  useful to observe the process of collapse of a state, and to engineer feedback to control its dynamics. In fact, it can be shown that a projective measurement is equivalent to an infinite number of continuous weak measurements carried out over an infinitesimally small time interval. Such a {continuous} measurement protocol leads to the conditional evolution of the density matrix based on noisy measurement outcomes given by,   
\begin{align}
\nonumber
d\rho_c(t) & =  - i [H, \rho_c(t)] dt + \kappa \mathcal{D}[A] \rho_c(t) dt \\
&+ \sqrt{\kappa} \mathcal{H}[A] \rho_c(t) dW(t), 
\label{eq:methods_sme}
\end{align}
where $\rho_c$ denotes the conditional density matrix of the system described by the Hamiltonian $H$. $A = \mathcal{A}/\mathcal{A}_0$ is a dimensionless operator corresponding to the physical observable $\mathcal{A}$ scaled suitably by $\mathcal{A}_0$, and is known as the measurement operator, which is measured with a measurement rate of $\kappa$ (denotes the rate at which the information is extracted). The first term of the above equation on the right-hand side represents the coherent evolution of the system. The second term on the right-hand side gives the measurement backaction due to the measurement of $A$, where $\mathcal{D}[A]\rho = A\rho A^\dagger - \frac{1}{2}(A^\dagger A \rho + \rho A^\dagger A)$ represents the decoherence superoperator. The last term is the stochastic diffusion term with $dW(t)$ being the Wiener increments, which are Gaussian distributed random variables with zero mean and represent  memoryless white noise, $\langle dW(t) dW(\tau)\rangle = \delta(t - \tau)$. $\mathcal H$ is a superoperator given by, $\mathcal{H}[A]\rho = A\rho + \rho A^\dagger - \rho \mathrm{tr}[A\rho + \rho A^\dagger]$. Eq.~\ref{eq:methods_sme} is known as the stochastic master equation (SME). The measurement records $dQ(t)$ are given by, 
\begin{align}
dQ(t) = \langle A_c (t) \rangle dt + \frac{1}{\sqrt{4\kappa}} {dW(t)},
\end{align}
where $\langle A_c (t) \rangle$ denotes the conditional mean of the measurement operator $A$ (dimensionless) at time $t$, which is nothing but the signal, and the last term represents the measurement noise associated with it.

Each evolution of the density matrix, $\rho_c(t)$ in time, following the SME in Eq.~\ref{eq:methods_sme}, represents a quantum trajectory, which can be manipulated and controlled by using appropriate feedback to the Hamiltonian in real time. If the feedback Hamiltonian, $F(t)$ is based on the conditional state, $\rho_c(t)$ or the conditional mean $\langle A_c (t) \rangle$ of the measurement operator, then the SME with the feedback Hamiltonian is given by, 
\begin{align}
\nonumber
d\rho_c(t) =&  - i [H, \rho_c(t)] dt + \kappa \mathcal{D}[A] \rho_c(t) dt \\
+& \sqrt{\kappa} \mathcal{H}[A] \rho_c(t) dW(t) - i [F(t), \rho_c(t)] dt. 
\end{align} 
For non-ideal measurement efficiency $\eta$ and in presence of the environmental decoherence, it becomes,
\begin{align}
\nonumber
d\rho_c(t) =&  - i [H, \rho_c(t)] dt + \gamma \mathcal{D}[c] \rho_c(t) dt  +\kappa \mathcal{D}[A] \rho_c(t) dt \\
&+\sqrt{\kappa \eta} \mathcal{H}[A] \rho_c(t) dW(t) - i [F(t), \rho_c(t)] dt,
\end{align}
where $\gamma$ is the environmental decoherence rate with collapse operator $c$. The expression for the continuous measurement record in presence of $\eta$ gets modified to, 
\begin{align}
dQ(t) = \langle A_c (t) \rangle dt + \frac{1}{\sqrt{4\eta\kappa }} {dW(t)}.
\end{align}

\textbf{Discrete quantum error correction.}
Generally speaking, QEC is a method to protect an unknown state of an open quantum system. However, in the context of quantum computing, we will consider qubits interacting with environmental decoherences. In contrast to classical bit errors there are two sources of errors, bit and phase flips. To correct bit-flip errors, stabilizer codes are used, while phase errors can be corrected similarly to the bit-flip errors but in a rotated basis (Hadamard basis) of the physical qubits~\cite{NielsenChuangBook}. Stabilizer codes are repetition codes, where the unknown state of the qubit is mapped onto a tensor space of a larger Hilbert space of multiple qubits as entangled states. Such an entangled unit of qubits is called a logical qubit. For example, in the three qubit repetition code, the unknown state of the qubit is mapped onto three physical qubits, on which single bit-flip errors can be corrected,
\begin{align}
    |0\rangle \to |000\rangle \equiv |0\rangle_L, \\
    |1\rangle \to |111\rangle \equiv |1\rangle_L.
\end{align}
Here the states $|0\rangle_L$ and $|1\rangle_L$ are the basis states for the QEC code and the space spanned by them is called the codespace. The elements of the codespace are known as the codewords. If the state of a physical qubit is $|\psi\rangle = \alpha|0\rangle + \beta|1\rangle$,  it is encoded to two more physical qubits as $|\psi\rangle_L = \alpha|000\rangle + \beta|111\rangle$, with $\alpha^2 + \beta^2 = 1$. The time evolution of the density matrix of the logical qubit under bit-flip errors caused by the environment decoherences, at a characteristic rate of $\gamma$ is described by,
\begin{align}
d\rho(t) = \gamma (\mathcal{D}[XII] + \mathcal{D}[IXI] + \mathcal{D}[IIX])\, \rho\, dt. 
\label{eq:methods_errors_dqec}
\end{align}
This is equivalent to assuming that the environment causes independent bit-flips of each physical qubit at Poisson distributed times with rate $\gamma$.

The essence of QEC is that the state of the logical qubit, $|\psi\rangle_L$ is  unknown to us, except however, the codespace, and we need to preserve it without losing the initial fidelity and  without any knowledge of the elements, $\alpha$ and $\beta$ of the state. In this situation, it is possible to measure a few special observables that determine the parities of the neighbouring qubits without giving any information about the state of the qubits themselves. In the three qubit code, there are three possible such operators, given by, $M_1 = ZZI$, $M_2 = IZZ$ and $M_3 = ZIZ$, where the third operator can be considered redundant. As $M_j^2=\mathbb{I}$, these operators have two possible eigenvalues $\pm 1$. The pair of eigenvalues $(m_1, m_2)$ for the simultaneous measurements of $M_1$ and $M_2$ gives the bit-flip error happening on a given qubit, provided no two qubits are flipped at the same time. Such one qubit flips can be corrected by applying unitary $X$ gates to the qubit on which the flip happened. Typically, to achieve this, the syndrome operators are projectively measured and errors are corrected based on the following conditions of the outcomes ($m_1, m_2$): (i) $(-1, +1) \to XII$, (ii) $(-1, -1) \to IXI$, (iii) $(+1, -1) \to IIX$ and (iv) $(-1, +1) \to $ None. We will call QEC based on projective measurement as discrete quantum error correction (DQEC) from now on. 

In order for DQEC to work, it is important to make the assumption that there are no multiple flips of the qubits happening simultaneously, and that no single flip errors are missed. Given the fact that projective measurements require significant time between each measurement, while the environment acts to degrade the qubits continuously, DQEC can never be conducted perfectly, and the error correction performance drops significantly over time. Theoretically speaking, if we consider each error to be  detected perfectly, the contribution of simultaneous bit-flips can be relatively small for low environmental decoherences, $\gamma$, as the theoretical fidelity of the error corrected logical state with DQEC with respect to the initial state is given by~\cite{Ahn2002Mar}, 
\begin{align}
F_{\rm DQEC}(t) = \frac{1}{4} ( 2 + 3e^{-2\gamma t} - e^{-6\gamma t}).
\end{align}
The drop in fidelity due to the bit-flip errors in a single qubit without error correction, is given by, 
\begin{align}
F_1(t) = \frac{1}{2} ( 1 + e^{-2\gamma t}),
\end{align}
and that of three qubits is given by $F_3(t) = F_1(t)^3$. This essentially means that $F_{\rm DQEC}(t) \sim F_1(t)$ when $t \geq \pi$, which shows how quickly the DQEC performance drops.

\textbf{Continuous quantum error correction.}
CQEC differs from DQEC in multiple aspects: in the way the measurements are performed on the syndrome operators, how the errors are detected, and how the errors are corrected. Instead of projective measurements, CQEC utilizes continuous weak measurements of the syndrome operators, discussed above. The conditional evolution of the state of the logical qubit undergoing bit-flip errors, continuous measurements and feedback is modelled using the SME as, 
\begin{align}
\nonumber
d\rho_c (t) = & \gamma (\mathcal D[XII] + \mathcal D [IXI] + \mathcal D [IIX]) \rho_c dt \\ \nonumber
+&  \kappa (\mathcal D[ZZI] + \mathcal D [IZZ] + \mathcal D [ZIZ]) \rho_c dt \\ \nonumber
+ & \sqrt \kappa (\mathcal H[ZZI] dW_1 + \mathcal H [IZZ] dW_2+ \mathcal H [ZIZ] dW_3) \rho_c \\
-&  i [F(t), \rho_c] dt, 
\end{align}
where the stochastic time varying measurement records of the stabilizer generators  are given by,
\begin{align}
\label{eq:current-zzi}
dQ_1(t) = &  \langle ZZI \rangle_c dt + \frac{1}{\sqrt{4\kappa }} {dW_1(t)},\\
\label{eq:current-izz}
dQ_2(t) = &  \langle IZZ \rangle_c dt + \frac{1}{\sqrt{4\kappa }} {dW_2(t)},\\
\label{eq:current-ziz}
dQ_3(t) = &  \langle ZIZ \rangle_c dt + \frac{1}{\sqrt{4\kappa }} {dW_3(t)}.
\end{align}
Here $F(t)$ is the feedback Hamiltonian given by,
\begin{align}
F(t) = \lambda_1(t) XII + \lambda_2(t) IXI + \lambda_3(t) IIX,
\end{align}
where $\lambda_i(t)$'s are, in principle, time dependent control parameters which depends on the conditional means of the error syndromes.  For example, the following feedback scheme was proposed by  Ahn et al.~\cite{Ahn2002Mar} for CQEC,
\begin{align}
\label{eqn:oldfb}
\lambda_1(t) =& \lambda(1 - \langle{ZZI}\rangle_c)
                             (1 + \langle {IZZ} \rangle_c)
                             (1 - \langle{ZIZ}\rangle_c),  \\
\lambda_2(t) =& \lambda(1 - \langle{ZZI}\rangle_c)
                             (1 - \langle{IZZ}\rangle_c)
                             (1 + \langle{ZIZ}\rangle_c),  \\
\lambda_3(t) =& \lambda(1 + \langle{ZZI}\rangle_c)
                             (1 - \langle{IZZ}\rangle_c)
                             (1 - \langle{ZIZ}\rangle_c),
\label{eqn:oldfblast}
\end{align}
where $\lambda$ is a feedback strength of the order of the measurement rate, $\kappa$.  The feedback function $F(t)$ described above, makes use of the conditional means of the syndrome operators, and thus, to perform CQEC ideally, we require detailed information about the time dependence of the conditional means of the syndrome generators. However, the conditional means are not available from the measurement records (Eq.~\ref{eq:current-zzi}-\ref{eq:current-ziz}) directly as these quantities are masked by  measurement noise  that is a fundamental component of all quantum measurements. The signal to noise ratio of such measurements can be typically quite poor. For practical purposes, researchers have previously used temporal filters to recover these conditional means from the noisy measurement records, with some filters possessing non-uniform temporal weights, biasing up the most recent records to avoid any lags or delays~\cite{Sarovar2004May,Atalaya2021Apr,Livingston2021Jul, Mabuchi2009Oct}. Of course real world devices already have limits on their response bandwidths. The effects of additional software/hardware filtering to smooth out the noisy measurement records will also degrade the signal or conditional means. However, if somehow we happen to know the conditional means in real time perfectly, the CQEC scheme would perform  optimally for bit-flip correction under the assumption that no two or more qubits flip simultaneously in a three qubit stabilizer code. In the following, we show how our proposed measurement based estimator (MBE) scheme allows us to achieve this. We call this MBE method of CQEC as MBE-CQEC scheme. 

\textbf{The MBE-CQEC scheme.}
We now describe a scheme that can perform faithful real-time estimation of any dynamical changes affecting the logical qubit. This MBE scheme will play a crucial role in detecting the bit-flip errors perfectly and therefore in applying the appropriate feedback $\lambda_j(t)$, in a manner, as we will show, that achieves ultra-high levels of protection of the unknown quantum state. Let us denote the laboratory-based quantum system that we wish to protect, as the $\cal R$eal system, and our $\cal E$stimator system is a numerical/computational model of the $\cal R$eal system, shown schematically in Fig.~1 in the main text. We consider that the internal dynamics of this $\cal R$eal system is described by the Hamiltonian $H^\mathcal{R} = H$, and its conditional density matrix is given by, $\rho^{\cal R}_c$, under continuous measurement via the measurement operator $A^\mathcal{R} = A$ (dimensionless). This can be described by the SME described above as, 
\begin{align}
\nonumber
d\rho^{\cal R}_c(t) = & - i [H, \rho^\mathcal{R}_c(t)] dt +  \gamma \mathcal{D}[c] \rho^{\cal R}_c(t) dt \\+& \kappa \mathcal{D}[A] \rho^{\cal R}_c(t) dt 
+ \sqrt{\kappa} \mathcal{H}[A] \rho^{\cal R}_c(t)\, dW^{\cal R}(t), 
\end{align}
where the superscript $\cal R$ is used to represent the $\cal R$eal lab system. The measurement record, $dQ^\mathcal{R}(t)$ is given by the summation of the conditional mean of the measurement operator and the corresponding random noise component of the measurement, 
\begin{align}
dQ^\mathcal{R}(t) = \langle A^\mathcal{R} (t) \rangle_c dt + \frac{1}{\sqrt{4\kappa}} {dW^\mathcal{R}(t)}. 
\end{align}
Now, we make an $\cal E$stimator ($\cal E$) of the $\cal R$eal on a computer with the same physical model ($H^\mathcal{E}=H^\mathcal{R} = H$), and the continuous measurement of the same observable ($A^\mathcal{E}=A^\mathcal{R} = A$), but start the $\cal E$stimator dynamics with a known initial state, $\rho^\mathcal{E}(0)$, which might be different from the initial state of the $\cal R$eal system. The dynamics of this $\cal E$stimator can be modelled as, 
\begin{align}
\nonumber
d\rho^{\cal E}_c(t) =&  - i [H, \rho^{\cal E}_c(t)] dt + \gamma \mathcal{D}[c] \rho^{\cal E}_c(t) dt + \kappa \mathcal{D}[A] \rho^{\cal E}_c(t) dt 
\\+& \sqrt{\kappa} \mathcal{H}[A] \rho^{\cal E}_c(t)\, dW^{\cal E}(t).
\end{align}
We now can {\it slave} the dynamics of this $\cal E$stimator model to the dynamics of the $\cal R$eal system by setting the $\cal E$stimator noise ${dW^\mathcal{E}(t)}$  as,
\begin{align}
 {dW^\mathcal{E}(t)} = {\sqrt{4\kappa}}\left[
dQ^\mathcal{R}(t)- \langle A^\mathcal{E} (t) \rangle_c dt \right],
\label{eq:dWe}
\end{align}
where the conditional mean $\langle A^\mathcal{E} (t) \rangle_c$, is obtained from the $\cal E$stimator, which is readily available without any extraneous noise.  Thus, the dynamics of the $\cal E$stimator follows the measurement records of the $\cal R$eal system as,
\begin{align}
\nonumber
d\rho^{\cal E}_c(t) =&  - i [H, \rho^{\cal E}_c(t)] dt +  \gamma \mathcal{D}[c] \rho^{\cal E}_c(t) dt + \kappa \mathcal{D}[A] \rho^{\cal E}_c(t) dt 
\\+& {2\kappa}\left[
dQ^R(t) - \langle A^{\cal E} (t) \rangle_c dt \right] \mathcal{H}[A] \rho^{\cal E}_c(t) .
\label{eq:mbe_scheme_main_eq}
\end{align}

Let us now consider that the $\cal R$eal system consists of a logical qubit comprising three physical qubits with an encoded unknown quantum state $|\psi^{\cal R}\rangle_L = |\psi^{\cal R}\rangle_L^{\alpha, \beta}=\alpha |000\rangle + \beta |111\rangle$, which we want to protect from bit-flip errors. The $\cal E$stimator is modelled similarly but with a different initial quantum state $|\psi^{\cal E}\rangle_L = |\psi^{\cal E}\rangle_L^{\alpha^\prime, \beta^\prime} = \alpha^\prime |000\rangle + \beta^\prime |111\rangle$, where $\alpha^\prime \neq \alpha$ and $\beta^\prime \neq \beta$. While the values of $\alpha^\prime$ and $\beta^\prime$ we can choose; $\alpha$ and $\beta$ for the $\cal R$eal system can be any possible values not known to us. 
The conditional mean in the $\cal R$eal system, is unknown as it is masked by the measurement noise as already stated. However, for the syndrome operators, being parity operators, the measurement signals (error syndromes) are independent of the coefficients ($\alpha$ and $\beta$) of the logical state $|\psi\rangle_L =\alpha |000\rangle + \beta |111\rangle$, but only depends on the codespace ($|000\rangle$ and $|111\rangle$). 
The unperturbed syndrome values, $\langle {\cal G}_i \rangle_c^{{\cal R}/{\cal E}}$ at $t=0$ satisfy,
\begin{align}
\langle {\cal G}_i \rangle_c^{\cal E}(0) =  \langle {\cal G}_i \rangle_c^{\cal R}(0)=1.
\label{eq:calG}
\end{align}
Here ${\cal G}_i$ represents the $i$\textsuperscript{th} stabilizer operator. Now using Eq.~\ref{eq:mbe_scheme_main_eq}, the $\cal E$stimator can be propagated to the next timestep after a measurement time interval of $dt$ using the measurement current from the $\cal R$eal but the conditional means from the $\cal E$stimator. 
For the second step, the noise signal can be correctly recovered as, $\langle {\cal G}_i \rangle_c^{\cal E}(dt) =  \langle {\cal G}_i \rangle_c^{\cal R}(dt)$, which can be either $+1$ or $-1$ unlike Eq.~\ref{eq:calG}, and similarly the process is repeated in timesteps of $dt$ for the $\cal E$stimator for subsequent times. Such an $\cal E$stimator that is  fed with real-time measurement records can correctly emulate the dynamics of all the errors happening on the $\cal R$eal system for each quantum trajectory. One can extract the error syndromes of the $\cal R$eal system by merely looking at the $\cal E$stimator conditional syndrome values, which are readily available. This solves the main problem of CQEC codes, where it is otherwise not possible to isolate the error syndromes from the measurement noise. The scheme is abbreviated as \textbf{MBE-CQEC} standing for \textit{measurement based estimator scheme for continuous quantum error correction}, and is shown schematically in Fig.~1 in the main text of the article.  

The MBE-CQEC scheme described above gives us a smart way of computing the error syndromes within a continuous measurement process, which allows us to correct bit-flips errors in real time in much rapid time intervals than DQEC codes~\cite{NielsenChuangBook} or using Eqn.~\ref{eqn:oldfb}-\ref{eqn:oldfblast}.  In the following we show how the estimator system has real time tomographic information about the errors happening to individual qubits, and we can use this to devise a new correction scheme.  This new scheme is not  based on the conditional means of the stabilizer operators, but instead, on the deviation of $\langle Z(t) \rangle$ of the qubits in the $\cal E$stimator relative to their original values at $t=0$, which is described below.

In the main text of the article, we have shown how the absolute deviation of the expectation value of the Pauli-$Z$ operator of the physical qubit $q$ of the $\cal E$stimator, $|\Delta \langle Z_q(t)\rangle = \langle Z_q(t) \rangle - \langle Z_q(0) \rangle|$  scaled by its initial unperturbed value $|\langle Z_q(0) \rangle|$, i.e.,  $|\Delta \langle Z_q(t)\rangle|/|\langle Z_q(0) \rangle|$, follows the same in the $\cal R$eal system. This constitutes the backbone of the error detection/correction proposal presented in the article. To understand it better, let's consider a physical qubit $q$ with state given by $|\psi\rangle = \alpha|0\rangle + \beta|1\rangle$. The initial expectation value $Z_q$ without measurement is $\langle Z_q(0) \rangle = \beta^2 - \alpha ^2$. A flip of the qubit at time $t > 0$ will mean $\langle Z_q(t) \rangle = \alpha^2 - \beta^2$, such that the absolute deviation from its initial state is given by, $|\langle \Delta Z_q(t) \rangle| = 2|\beta^2 - \alpha^2|$. The ratio $|\Delta \langle  Z_q(t) \rangle|/|\langle Z_q(0) \rangle| = 2$, which means a complete flip. Similarly, $|\Delta  \langle  Z_q(t) \rangle|/| \langle  Z_q(0) \rangle|=0$ would mean absolutely no flipping. For any other change $\varepsilon$ in between, $|\Delta  \langle  Z_q(t) \rangle|/| \langle  Z_q(0) \rangle|=\varepsilon$. The $\cal E$stimator qubit can be modelled with $\alpha = 1$ and $\beta = 0$, i.e., at the reset condition for convenience, which means $\langle Z_q(0) \rangle=1$ for the $\cal E$stimator qubit. Under the same noise measurement signals, $dW_{s}^{\cal E}(t) = dW_{s}^{\cal E}$, where $s = (ZZI, IZZ, ZIZ)$ denote the syndrome operators under measurement of the three physical qubits of the $\cal R$eal system, a change in $|\Delta  \langle  Z^{\mathcal R}_q(t) \rangle|/| \langle  Z^{\mathcal R}_q(0) \rangle|$ on qubit $q$ by an amount $\varepsilon$ will underpin similar changes in the $\cal E$stimator qubits, $|\Delta  \langle  Z^{\mathcal E}_q(t) \rangle(t)|/| \langle  Z^{\mathcal E}_q(0) \rangle|$, i.e.,
\begin{align}
\frac{|\Delta  \langle  Z^{\mathcal E}_q(t) \rangle|}{| \langle  Z^{\mathcal E}_q(0) \rangle|} = \frac{|\Delta  \langle  Z^{\mathcal R}_q(t) \rangle|}{| \langle  Z^{\mathcal R}_q(0) \rangle|} = \varepsilon(t).
\end{align}
Thus, we can use the following condition on the $\cal E$stimator system to detect a bit-flip error on qubit $q$ and correspondingly apply the feedback 
Hamiltonian after a time $\delta t$,
\begin{equation}
\lambda_q ( t + \delta t) = 
\begin{cases}
    \lambda_q, &\mathrm{if}|\langle Z_q^{\cal E}(t) \rangle - \langle Z_q^{\cal E}(0) \rangle| > \varepsilon |\langle Z_q^{\cal E}(0) \rangle|,\\
    0,              & \text{otherwise},
\end{cases}
\end{equation}
where $\lambda_q \sim \kappa$ and $\epsilon$ is a tolerance slightly higher than 1, which we choose to be 
$\varepsilon = 1.05$.

\section{acknowledgments}
	The authors thank the super-computing facilities provided by the Okinawa Institute of Science and Technology (OIST) Graduate University and financial support. GJM acknowledge the support of the Australian Research Council Centre of Excellence for Engineered Quantum Systems CE170100009. 

% \section{Author contribution}
% The project was conceived by SB in discussion with BS and JT. SB conducted the study and prepared the first draft of the manuscript. All authors participated in regular discussions of the results and interpretations that contributed to better understanding and improvement. All authors contributed in the process of getting the manuscript completed in publishable form.  

% \section{Data availability}
% Data supporting the findings of this study are available within the article and
% its Supplementary Information and from the corresponding author upon reasonable request.

% TC:ignore
% \bibliographystyle{apsrev4-1}
% \bibliography{bibfile}

\begin{thebibliography}{36}%
\makeatletter
\providecommand \@ifxundefined [1]{%
 \@ifx{#1\undefined}
}%
\providecommand \@ifnum [1]{%
 \ifnum #1\expandafter \@firstoftwo
 \else \expandafter \@secondoftwo
 \fi
}%
\providecommand \@ifx [1]{%
 \ifx #1\expandafter \@firstoftwo
 \else \expandafter \@secondoftwo
 \fi
}%
\providecommand \natexlab [1]{#1}%
\providecommand \enquote  [1]{``#1''}%
\providecommand \bibnamefont  [1]{#1}%
\providecommand \bibfnamefont [1]{#1}%
\providecommand \citenamefont [1]{#1}%
\providecommand \href@noop [0]{\@secondoftwo}%
\providecommand \href [0]{\begingroup \@sanitize@url \@href}%
\providecommand \@href[1]{\@@startlink{#1}\@@href}%
\providecommand \@@href[1]{\endgroup#1\@@endlink}%
\providecommand \@sanitize@url [0]{\catcode `\\12\catcode `\$12\catcode
  `\&12\catcode `\#12\catcode `\^12\catcode `\_12\catcode `\%12\relax}%
\providecommand \@@startlink[1]{}%
\providecommand \@@endlink[0]{}%
\providecommand \url  [0]{\begingroup\@sanitize@url \@url }%
\providecommand \@url [1]{\endgroup\@href {#1}{\urlprefix }}%
\providecommand \urlprefix  [0]{URL }%
\providecommand \Eprint [0]{\href }%
\providecommand \doibase [0]{http://dx.doi.org/}%
\providecommand \selectlanguage [0]{\@gobble}%
\providecommand \bibinfo  [0]{\@secondoftwo}%
\providecommand \bibfield  [0]{\@secondoftwo}%
\providecommand \translation [1]{[#1]}%
\providecommand \BibitemOpen [0]{}%
\providecommand \bibitemStop [0]{}%
\providecommand \bibitemNoStop [0]{.\EOS\space}%
\providecommand \EOS [0]{\spacefactor3000\relax}%
\providecommand \BibitemShut  [1]{\csname bibitem#1\endcsname}%
\let\auto@bib@innerbib\@empty
%</preamble>
\bibitem [{\citenamefont {Nielsen}\ and\ \citenamefont
  {Chuang}(2010)}]{NielsenChuangBook}%
  \BibitemOpen
  \bibfield  {author} {\bibinfo {author} {\bibfnamefont {Michael~A.}\
  \bibnamefont {Nielsen}}\ and\ \bibinfo {author} {\bibfnamefont {Isaac~L.}\
  \bibnamefont {Chuang}},\ }\href {\doibase 10.1017/CBO9780511976667} {\emph
  {\bibinfo {title} {{Quantum Computation and Quantum Information}}}}\
  (\bibinfo  {publisher} {Cambridge University Press},\ \bibinfo {address}
  {Cambridge, England, UK},\ \bibinfo {year} {2010})\BibitemShut {NoStop}%
\bibitem [{\citenamefont {Shor}(1995)}]{Shor1995Oct}%
  \BibitemOpen
  \bibfield  {author} {\bibinfo {author} {\bibfnamefont {Peter~W.}\
  \bibnamefont {Shor}},\ }\bibfield  {title} {\enquote {\bibinfo {title}
  {{Scheme for reducing decoherence in quantum computer memory}},}\ }\href
  {\doibase 10.1103/PhysRevA.52.R2493} {\bibfield  {journal} {\bibinfo
  {journal} {Phys. Rev. A}\ }\textbf {\bibinfo {volume} {52}},\ \bibinfo
  {pages} {R2493--R2496(R)} (\bibinfo {year} {1995})}\BibitemShut {NoStop}%
\bibitem [{\citenamefont {Steane}(1996)}]{Steane1996Jul}%
  \BibitemOpen
  \bibfield  {author} {\bibinfo {author} {\bibfnamefont {A.~M.}\ \bibnamefont
  {Steane}},\ }\bibfield  {title} {\enquote {\bibinfo {title} {{Error
  Correcting Codes in Quantum Theory}},}\ }\href {\doibase
  10.1103/PhysRevLett.77.793} {\bibfield  {journal} {\bibinfo  {journal} {Phys.
  Rev. Lett.}\ }\textbf {\bibinfo {volume} {77}},\ \bibinfo {pages} {793--797}
  (\bibinfo {year} {1996})}\BibitemShut {NoStop}%
\bibitem [{\citenamefont {Djordjevic}(2012)}]{Djordjevic2012}%
  \BibitemOpen
  \bibfield  {author} {\bibinfo {author} {\bibfnamefont {Ivan}\ \bibnamefont
  {Djordjevic}},\ }\href {\doibase 10.1016/C2010-0-66917-3} {\emph {\bibinfo
  {title} {{Quantum Information Processing and Quantum Error Correction}}}}\
  (\bibinfo  {publisher} {Elsevier, Academic Press},\ \bibinfo {year}
  {2012})\BibitemShut {NoStop}%
\bibitem [{\citenamefont {Gertler}\ \emph {et~al.}(2021)\citenamefont
  {Gertler}, \citenamefont {Baker}, \citenamefont {Li}, \citenamefont {Shirol},
  \citenamefont {Koch},\ and\ \citenamefont {Wang}}]{Gertler2021Feb}%
  \BibitemOpen
  \bibfield  {author} {\bibinfo {author} {\bibfnamefont {Jeffrey~M.}\
  \bibnamefont {Gertler}}, \bibinfo {author} {\bibfnamefont {Brian}\
  \bibnamefont {Baker}}, \bibinfo {author} {\bibfnamefont {Juliang}\
  \bibnamefont {Li}}, \bibinfo {author} {\bibfnamefont {Shruti}\ \bibnamefont
  {Shirol}}, \bibinfo {author} {\bibfnamefont {Jens}\ \bibnamefont {Koch}}, \
  and\ \bibinfo {author} {\bibfnamefont {Chen}\ \bibnamefont {Wang}},\
  }\bibfield  {title} {\enquote {\bibinfo {title} {{Protecting a bosonic qubit
  with autonomous quantum error correction}},}\ }\href {\doibase
  10.1038/s41586-021-03257-0} {\bibfield  {journal} {\bibinfo  {journal}
  {Nature}\ }\textbf {\bibinfo {volume} {590}},\ \bibinfo {pages} {243--248}
  (\bibinfo {year} {2021})}\BibitemShut {NoStop}%
\bibitem [{\citenamefont {Gottesman}(1997)}]{Gottesman1997May}%
  \BibitemOpen
  \bibfield  {author} {\bibinfo {author} {\bibfnamefont {Daniel}\ \bibnamefont
  {Gottesman}},\ }\bibfield  {title} {\enquote {\bibinfo {title} {{Stabilizer
  Codes and Quantum Error Correction}},}\ }\href
  {https://arxiv.org/abs/quant-ph/9705052v1} {\bibfield  {journal} {\bibinfo
  {journal} {arXiv}\ } (\bibinfo {year} {1997})},\ \Eprint
  {http://arxiv.org/abs/quant-ph/9705052} {quant-ph/9705052} \BibitemShut
  {NoStop}%
\bibitem [{\citenamefont {Devitt}\ \emph {et~al.}(2013)\citenamefont {Devitt},
  \citenamefont {Munro},\ and\ \citenamefont {Nemoto}}]{Devitt2013Jun}%
  \BibitemOpen
  \bibfield  {author} {\bibinfo {author} {\bibfnamefont {Simon~J.}\
  \bibnamefont {Devitt}}, \bibinfo {author} {\bibfnamefont {William~J.}\
  \bibnamefont {Munro}}, \ and\ \bibinfo {author} {\bibfnamefont {Kae}\
  \bibnamefont {Nemoto}},\ }\bibfield  {title} {\enquote {\bibinfo {title}
  {{Quantum error correction for beginners}},}\ }\href {\doibase
  10.1088/0034-4885/76/7/076001} {\bibfield  {journal} {\bibinfo  {journal}
  {Rep. Prog. Phys.}\ }\textbf {\bibinfo {volume} {76}},\ \bibinfo {pages}
  {076001} (\bibinfo {year} {2013})}\BibitemShut {NoStop}%
\bibitem [{\citenamefont {Reed}\ \emph {et~al.}(2012)\citenamefont {Reed},
  \citenamefont {DiCarlo}, \citenamefont {Nigg}, \citenamefont {Sun},
  \citenamefont {Frunzio}, \citenamefont {Girvin},\ and\ \citenamefont
  {Schoelkopf}}]{Reed2012Feb}%
  \BibitemOpen
  \bibfield  {author} {\bibinfo {author} {\bibfnamefont {M.~D.}\ \bibnamefont
  {Reed}}, \bibinfo {author} {\bibfnamefont {L.}~\bibnamefont {DiCarlo}},
  \bibinfo {author} {\bibfnamefont {S.~E.}\ \bibnamefont {Nigg}}, \bibinfo
  {author} {\bibfnamefont {L.}~\bibnamefont {Sun}}, \bibinfo {author}
  {\bibfnamefont {L.}~\bibnamefont {Frunzio}}, \bibinfo {author} {\bibfnamefont
  {S.~M.}\ \bibnamefont {Girvin}}, \ and\ \bibinfo {author} {\bibfnamefont
  {R.~J.}\ \bibnamefont {Schoelkopf}},\ }\bibfield  {title} {\enquote {\bibinfo
  {title} {{Realization of three-qubit quantum error correction with
  superconducting circuits}},}\ }\href {\doibase 10.1038/nature10786}
  {\bibfield  {journal} {\bibinfo  {journal} {Nature}\ }\textbf {\bibinfo
  {volume} {482}},\ \bibinfo {pages} {382--385} (\bibinfo {year}
  {2012})}\BibitemShut {NoStop}%
\bibitem [{\citenamefont {Knill}\ \emph {et~al.}(1998)\citenamefont {Knill},
  \citenamefont {Laflamme},\ and\ \citenamefont {Zurek}}]{Knill1998Jan}%
  \BibitemOpen
  \bibfield  {author} {\bibinfo {author} {\bibfnamefont {Emanuel}\ \bibnamefont
  {Knill}}, \bibinfo {author} {\bibfnamefont {Raymond}\ \bibnamefont
  {Laflamme}}, \ and\ \bibinfo {author} {\bibfnamefont {Wojciech~H.}\
  \bibnamefont {Zurek}},\ }\bibfield  {title} {\enquote {\bibinfo {title}
  {{Resilient quantum computation: error models and thresholds}},}\ }\href
  {\doibase 10.1098/rspa.1998.0166} {\bibfield  {journal} {\bibinfo  {journal}
  {Proc. R. Soc. Lond. A.}\ }\textbf {\bibinfo {volume} {454}},\ \bibinfo
  {pages} {365--384} (\bibinfo {year} {1998})}\BibitemShut {NoStop}%
\bibitem [{\citenamefont {Girvin}(2021)}]{Girvin2021Nov}%
  \BibitemOpen
  \bibfield  {author} {\bibinfo {author} {\bibfnamefont {Steven~M.}\
  \bibnamefont {Girvin}},\ }\bibfield  {title} {\enquote {\bibinfo {title}
  {{Introduction to Quantum Error Correction and Fault Tolerance}},}\ }\href
  {https://arxiv.org/abs/2111.08894v1} {\bibfield  {journal} {\bibinfo
  {journal} {arXiv}\ } (\bibinfo {year} {2021})},\ \Eprint
  {http://arxiv.org/abs/2111.08894} {2111.08894} \BibitemShut {NoStop}%
\bibitem [{\citenamefont {Schindler}\ \emph {et~al.}(2011)\citenamefont
  {Schindler}, \citenamefont {Barreiro}, \citenamefont {Monz}, \citenamefont
  {Nebendahl}, \citenamefont {Nigg}, \citenamefont {Chwalla}, \citenamefont
  {Hennrich},\ and\ \citenamefont {Blatt}}]{Schindler2011May}%
  \BibitemOpen
  \bibfield  {author} {\bibinfo {author} {\bibfnamefont {Philipp}\ \bibnamefont
  {Schindler}}, \bibinfo {author} {\bibfnamefont {Julio~T.}\ \bibnamefont
  {Barreiro}}, \bibinfo {author} {\bibfnamefont {Thomas}\ \bibnamefont {Monz}},
  \bibinfo {author} {\bibfnamefont {Volckmar}\ \bibnamefont {Nebendahl}},
  \bibinfo {author} {\bibfnamefont {Daniel}\ \bibnamefont {Nigg}}, \bibinfo
  {author} {\bibfnamefont {Michael}\ \bibnamefont {Chwalla}}, \bibinfo {author}
  {\bibfnamefont {Markus}\ \bibnamefont {Hennrich}}, \ and\ \bibinfo {author}
  {\bibfnamefont {Rainer}\ \bibnamefont {Blatt}},\ }\bibfield  {title}
  {\enquote {\bibinfo {title} {{Experimental Repetitive Quantum Error
  Correction}},}\ }\href {\doibase 10.1126/science.1203329} {\bibfield
  {journal} {\bibinfo  {journal} {Science}\ }\textbf {\bibinfo {volume}
  {332}},\ \bibinfo {pages} {1059--1061} (\bibinfo {year} {2011})}\BibitemShut
  {NoStop}%
\bibitem [{\citenamefont {Linke}\ \emph {et~al.}(2017)\citenamefont {Linke},
  \citenamefont {Gutierrez}, \citenamefont {Landsman}, \citenamefont {Figgatt},
  \citenamefont {Debnath}, \citenamefont {Brown},\ and\ \citenamefont
  {Monroe}}]{Linke2017Oct}%
  \BibitemOpen
  \bibfield  {author} {\bibinfo {author} {\bibfnamefont {Norbert~M.}\
  \bibnamefont {Linke}}, \bibinfo {author} {\bibfnamefont {Mauricio}\
  \bibnamefont {Gutierrez}}, \bibinfo {author} {\bibfnamefont {Kevin~A.}\
  \bibnamefont {Landsman}}, \bibinfo {author} {\bibfnamefont {Caroline}\
  \bibnamefont {Figgatt}}, \bibinfo {author} {\bibfnamefont {Shantanu}\
  \bibnamefont {Debnath}}, \bibinfo {author} {\bibfnamefont {Kenneth~R.}\
  \bibnamefont {Brown}}, \ and\ \bibinfo {author} {\bibfnamefont {Christopher}\
  \bibnamefont {Monroe}},\ }\bibfield  {title} {\enquote {\bibinfo {title}
  {{Fault-tolerant quantum error detection}},}\ }\href {\doibase
  10.1126/sciadv.1701074} {\bibfield  {journal} {\bibinfo  {journal} {Sci.
  Adv.}\ }\textbf {\bibinfo {volume} {3}},\ \bibinfo {pages} {e1701074}
  (\bibinfo {year} {2017})}\BibitemShut {NoStop}%
\bibitem [{\citenamefont {Negnevitsky}\ \emph {et~al.}(2018)\citenamefont
  {Negnevitsky}, \citenamefont {Marinelli}, \citenamefont {Mehta},
  \citenamefont {Lo}, \citenamefont {Fl{\ifmmode\ddot{u}\else\"{u}\fi}hmann},\
  and\ \citenamefont {Home}}]{Negnevitsky2018Nov}%
  \BibitemOpen
  \bibfield  {author} {\bibinfo {author} {\bibfnamefont {V.}~\bibnamefont
  {Negnevitsky}}, \bibinfo {author} {\bibfnamefont {M.}~\bibnamefont
  {Marinelli}}, \bibinfo {author} {\bibfnamefont {K.~K.}\ \bibnamefont
  {Mehta}}, \bibinfo {author} {\bibfnamefont {H.-Y.}\ \bibnamefont {Lo}},
  \bibinfo {author} {\bibfnamefont {C.}~\bibnamefont
  {Fl{\ifmmode\ddot{u}\else\"{u}\fi}hmann}}, \ and\ \bibinfo {author}
  {\bibfnamefont {J.~P.}\ \bibnamefont {Home}},\ }\bibfield  {title} {\enquote
  {\bibinfo {title} {{Repeated multi-qubit readout and feedback with a
  mixed-species trapped-ion register}},}\ }\href {\doibase
  10.1038/s41586-018-0668-z} {\bibfield  {journal} {\bibinfo  {journal}
  {Nature}\ }\textbf {\bibinfo {volume} {563}},\ \bibinfo {pages} {527--531}
  (\bibinfo {year} {2018})}\BibitemShut {NoStop}%
\bibitem [{\citenamefont {Cramer}\ \emph {et~al.}(2016)\citenamefont {Cramer},
  \citenamefont {Kalb}, \citenamefont {Rol}, \citenamefont {Hensen},
  \citenamefont {Blok}, \citenamefont {Markham}, \citenamefont {Twitchen},
  \citenamefont {Hanson},\ and\ \citenamefont {Taminiau}}]{Cramer2016May}%
  \BibitemOpen
  \bibfield  {author} {\bibinfo {author} {\bibfnamefont {J.}~\bibnamefont
  {Cramer}}, \bibinfo {author} {\bibfnamefont {N.}~\bibnamefont {Kalb}},
  \bibinfo {author} {\bibfnamefont {M.~A.}\ \bibnamefont {Rol}}, \bibinfo
  {author} {\bibfnamefont {B.}~\bibnamefont {Hensen}}, \bibinfo {author}
  {\bibfnamefont {M.~S.}\ \bibnamefont {Blok}}, \bibinfo {author}
  {\bibfnamefont {M.}~\bibnamefont {Markham}}, \bibinfo {author} {\bibfnamefont
  {D.~J.}\ \bibnamefont {Twitchen}}, \bibinfo {author} {\bibfnamefont
  {R.}~\bibnamefont {Hanson}}, \ and\ \bibinfo {author} {\bibfnamefont {T.~H.}\
  \bibnamefont {Taminiau}},\ }\bibfield  {title} {\enquote {\bibinfo {title}
  {{Repeated quantum error correction on a continuously encoded qubit by
  real-time feedback}},}\ }\href {\doibase 10.1038/ncomms11526} {\bibfield
  {journal} {\bibinfo  {journal} {Nat. Commun.}\ }\textbf {\bibinfo {volume}
  {7}},\ \bibinfo {pages} {1--7} (\bibinfo {year} {2016})}\BibitemShut
  {NoStop}%
\bibitem [{\citenamefont {Kelly}\ \emph {et~al.}(2015)\citenamefont {Kelly},
  \citenamefont {Barends}, \citenamefont {Fowler}, \citenamefont {Megrant},
  \citenamefont {Jeffrey}, \citenamefont {White}, \citenamefont {Sank},
  \citenamefont {Mutus}, \citenamefont {Campbell}, \citenamefont {Chen},
  \citenamefont {Chen}, \citenamefont {Chiaro}, \citenamefont {Dunsworth},
  \citenamefont {Hoi}, \citenamefont {Neill}, \citenamefont {O{'}Malley},
  \citenamefont {Quintana}, \citenamefont {Roushan}, \citenamefont
  {Vainsencher}, \citenamefont {Wenner}, \citenamefont {Cleland},\ and\
  \citenamefont {Martinis}}]{Kelly2015Mar}%
  \BibitemOpen
  \bibfield  {author} {\bibinfo {author} {\bibfnamefont {J.}~\bibnamefont
  {Kelly}}, \bibinfo {author} {\bibfnamefont {R.}~\bibnamefont {Barends}},
  \bibinfo {author} {\bibfnamefont {A.~G.}\ \bibnamefont {Fowler}}, \bibinfo
  {author} {\bibfnamefont {A.}~\bibnamefont {Megrant}}, \bibinfo {author}
  {\bibfnamefont {E.}~\bibnamefont {Jeffrey}}, \bibinfo {author} {\bibfnamefont
  {T.~C.}\ \bibnamefont {White}}, \bibinfo {author} {\bibfnamefont
  {D.}~\bibnamefont {Sank}}, \bibinfo {author} {\bibfnamefont {J.~Y.}\
  \bibnamefont {Mutus}}, \bibinfo {author} {\bibfnamefont {B.}~\bibnamefont
  {Campbell}}, \bibinfo {author} {\bibfnamefont {Yu}~\bibnamefont {Chen}},
  \bibinfo {author} {\bibfnamefont {Z.}~\bibnamefont {Chen}}, \bibinfo {author}
  {\bibfnamefont {B.}~\bibnamefont {Chiaro}}, \bibinfo {author} {\bibfnamefont
  {A.}~\bibnamefont {Dunsworth}}, \bibinfo {author} {\bibfnamefont {I.-C.}\
  \bibnamefont {Hoi}}, \bibinfo {author} {\bibfnamefont {C.}~\bibnamefont
  {Neill}}, \bibinfo {author} {\bibfnamefont {P.~J.~J.}\ \bibnamefont
  {O{'}Malley}}, \bibinfo {author} {\bibfnamefont {C.}~\bibnamefont
  {Quintana}}, \bibinfo {author} {\bibfnamefont {P.}~\bibnamefont {Roushan}},
  \bibinfo {author} {\bibfnamefont {A.}~\bibnamefont {Vainsencher}}, \bibinfo
  {author} {\bibfnamefont {J.}~\bibnamefont {Wenner}}, \bibinfo {author}
  {\bibfnamefont {A.~N.}\ \bibnamefont {Cleland}}, \ and\ \bibinfo {author}
  {\bibfnamefont {John~M.}\ \bibnamefont {Martinis}},\ }\bibfield  {title}
  {\enquote {\bibinfo {title} {{State preservation by repetitive error
  detection in a superconducting quantum circuit}},}\ }\href {\doibase
  10.1038/nature14270} {\bibfield  {journal} {\bibinfo  {journal} {Nature}\
  }\textbf {\bibinfo {volume} {519}},\ \bibinfo {pages} {66--69} (\bibinfo
  {year} {2015})}\BibitemShut {NoStop}%
\bibitem [{\citenamefont {Rist{\ifmmode\grave{e}\else\`{e}\fi}}\ \emph
  {et~al.}(2015)\citenamefont {Rist{\ifmmode\grave{e}\else\`{e}\fi}},
  \citenamefont {Poletto}, \citenamefont {Huang}, \citenamefont {Bruno},
  \citenamefont {Vesterinen}, \citenamefont {Saira},\ and\ \citenamefont
  {DiCarlo}}]{Riste2015Apr}%
  \BibitemOpen
  \bibfield  {author} {\bibinfo {author} {\bibfnamefont {D.}~\bibnamefont
  {Rist{\ifmmode\grave{e}\else\`{e}\fi}}}, \bibinfo {author} {\bibfnamefont
  {S.}~\bibnamefont {Poletto}}, \bibinfo {author} {\bibfnamefont {M.-Z.}\
  \bibnamefont {Huang}}, \bibinfo {author} {\bibfnamefont {A.}~\bibnamefont
  {Bruno}}, \bibinfo {author} {\bibfnamefont {V.}~\bibnamefont {Vesterinen}},
  \bibinfo {author} {\bibfnamefont {O.-P.}\ \bibnamefont {Saira}}, \ and\
  \bibinfo {author} {\bibfnamefont {L.}~\bibnamefont {DiCarlo}},\ }\bibfield
  {title} {\enquote {\bibinfo {title} {{Detecting bit-flip errors in a logical
  qubit using stabilizer measurements}},}\ }\href {\doibase 10.1038/ncomms7983}
  {\bibfield  {journal} {\bibinfo  {journal} {Nat. Commun.}\ }\textbf {\bibinfo
  {volume} {6}},\ \bibinfo {pages} {1--6} (\bibinfo {year} {2015})}\BibitemShut
  {NoStop}%
\bibitem [{\citenamefont {Ofek}\ \emph {et~al.}(2016)\citenamefont {Ofek},
  \citenamefont {Petrenko}, \citenamefont {Heeres}, \citenamefont {Reinhold},
  \citenamefont {Leghtas}, \citenamefont {Vlastakis}, \citenamefont {Liu},
  \citenamefont {Frunzio}, \citenamefont {Girvin}, \citenamefont {Jiang},
  \citenamefont {Mirrahimi}, \citenamefont {Devoret},\ and\ \citenamefont
  {Schoelkopf}}]{Ofek2016Aug}%
  \BibitemOpen
  \bibfield  {author} {\bibinfo {author} {\bibfnamefont {Nissim}\ \bibnamefont
  {Ofek}}, \bibinfo {author} {\bibfnamefont {Andrei}\ \bibnamefont {Petrenko}},
  \bibinfo {author} {\bibfnamefont {Reinier}\ \bibnamefont {Heeres}}, \bibinfo
  {author} {\bibfnamefont {Philip}\ \bibnamefont {Reinhold}}, \bibinfo {author}
  {\bibfnamefont {Zaki}\ \bibnamefont {Leghtas}}, \bibinfo {author}
  {\bibfnamefont {Brian}\ \bibnamefont {Vlastakis}}, \bibinfo {author}
  {\bibfnamefont {Yehan}\ \bibnamefont {Liu}}, \bibinfo {author} {\bibfnamefont
  {Luigi}\ \bibnamefont {Frunzio}}, \bibinfo {author} {\bibfnamefont {S.~M.}\
  \bibnamefont {Girvin}}, \bibinfo {author} {\bibfnamefont {L.}~\bibnamefont
  {Jiang}}, \bibinfo {author} {\bibfnamefont {Mazyar}\ \bibnamefont
  {Mirrahimi}}, \bibinfo {author} {\bibfnamefont {M.~H.}\ \bibnamefont
  {Devoret}}, \ and\ \bibinfo {author} {\bibfnamefont {R.~J.}\ \bibnamefont
  {Schoelkopf}},\ }\bibfield  {title} {\enquote {\bibinfo {title} {{Extending
  the lifetime of a quantum bit with error correction in superconducting
  circuits}},}\ }\href {\doibase 10.1038/nature18949} {\bibfield  {journal}
  {\bibinfo  {journal} {Nature}\ }\textbf {\bibinfo {volume} {536}},\ \bibinfo
  {pages} {441--445} (\bibinfo {year} {2016})}\BibitemShut {NoStop}%
\bibitem [{\citenamefont {{Arute et al.}}(2019)}]{Arute2019Oct}%
  \BibitemOpen
  \bibfield  {author} {\bibinfo {author} {\bibfnamefont {Frank}\ \bibnamefont
  {{Arute et al.}}},\ }\bibfield  {title} {\enquote {\bibinfo {title} {{Quantum
  supremacy using a programmable superconducting processor}},}\ }\href
  {\doibase 10.1038/s41586-019-1666-5} {\bibfield  {journal} {\bibinfo
  {journal} {Nature}\ }\textbf {\bibinfo {volume} {574}},\ \bibinfo {pages}
  {505--510} (\bibinfo {year} {2019})}\BibitemShut {NoStop}%
\bibitem [{\citenamefont {Andersen}\ \emph {et~al.}(2020)\citenamefont
  {Andersen}, \citenamefont {Remm}, \citenamefont {Lazar}, \citenamefont
  {Krinner}, \citenamefont {Lacroix}, \citenamefont {Norris}, \citenamefont
  {Gabureac}, \citenamefont {Eichler},\ and\ \citenamefont
  {Wallraff}}]{Andersen2020Aug}%
  \BibitemOpen
  \bibfield  {author} {\bibinfo {author} {\bibfnamefont {Christian~Kraglund}\
  \bibnamefont {Andersen}}, \bibinfo {author} {\bibfnamefont {Ants}\
  \bibnamefont {Remm}}, \bibinfo {author} {\bibfnamefont {Stefania}\
  \bibnamefont {Lazar}}, \bibinfo {author} {\bibfnamefont {Sebastian}\
  \bibnamefont {Krinner}}, \bibinfo {author} {\bibfnamefont {Nathan}\
  \bibnamefont {Lacroix}}, \bibinfo {author} {\bibfnamefont {Graham~J.}\
  \bibnamefont {Norris}}, \bibinfo {author} {\bibfnamefont {Mihai}\
  \bibnamefont {Gabureac}}, \bibinfo {author} {\bibfnamefont {Christopher}\
  \bibnamefont {Eichler}}, \ and\ \bibinfo {author} {\bibfnamefont {Andreas}\
  \bibnamefont {Wallraff}},\ }\bibfield  {title} {\enquote {\bibinfo {title}
  {{Repeated quantum error detection in a surface code}},}\ }\href {\doibase
  10.1038/s41567-020-0920-y} {\bibfield  {journal} {\bibinfo  {journal} {Nat.
  Phys.}\ }\textbf {\bibinfo {volume} {16}},\ \bibinfo {pages} {875--880}
  (\bibinfo {year} {2020})}\BibitemShut {NoStop}%
\bibitem [{\citenamefont {Stricker}\ \emph {et~al.}(2020)\citenamefont
  {Stricker}, \citenamefont {Vodola}, \citenamefont {Erhard}, \citenamefont
  {Postler}, \citenamefont {Meth}, \citenamefont {Ringbauer}, \citenamefont
  {Schindler}, \citenamefont {Monz}, \citenamefont
  {M{\ifmmode\ddot{u}\else\"{u}\fi}ller},\ and\ \citenamefont
  {Blatt}}]{Stricker2020Sep}%
  \BibitemOpen
  \bibfield  {author} {\bibinfo {author} {\bibfnamefont {Roman}\ \bibnamefont
  {Stricker}}, \bibinfo {author} {\bibfnamefont {Davide}\ \bibnamefont
  {Vodola}}, \bibinfo {author} {\bibfnamefont {Alexander}\ \bibnamefont
  {Erhard}}, \bibinfo {author} {\bibfnamefont {Lukas}\ \bibnamefont {Postler}},
  \bibinfo {author} {\bibfnamefont {Michael}\ \bibnamefont {Meth}}, \bibinfo
  {author} {\bibfnamefont {Martin}\ \bibnamefont {Ringbauer}}, \bibinfo
  {author} {\bibfnamefont {Philipp}\ \bibnamefont {Schindler}}, \bibinfo
  {author} {\bibfnamefont {Thomas}\ \bibnamefont {Monz}}, \bibinfo {author}
  {\bibfnamefont {Markus}\ \bibnamefont
  {M{\ifmmode\ddot{u}\else\"{u}\fi}ller}}, \ and\ \bibinfo {author}
  {\bibfnamefont {Rainer}\ \bibnamefont {Blatt}},\ }\bibfield  {title}
  {\enquote {\bibinfo {title} {{Experimental deterministic correction of qubit
  loss}},}\ }\href {\doibase 10.1038/s41586-020-2667-0} {\bibfield  {journal}
  {\bibinfo  {journal} {Nature}\ }\textbf {\bibinfo {volume} {585}},\ \bibinfo
  {pages} {207--210} (\bibinfo {year} {2020})}\BibitemShut {NoStop}%
\bibitem [{\citenamefont {{Chen et al.}}(2021)}]{Chen2021Jul}%
  \BibitemOpen
  \bibfield  {author} {\bibinfo {author} {\bibfnamefont {Zijun}\ \bibnamefont
  {{Chen et al.}}},\ }\bibfield  {title} {\enquote {\bibinfo {title}
  {{Exponential suppression of bit or phase errors with cyclic error
  correction}},}\ }\href {\doibase 10.1038/s41586-021-03588-y} {\bibfield
  {journal} {\bibinfo  {journal} {Nature}\ }\textbf {\bibinfo {volume} {595}},\
  \bibinfo {pages} {383--387} (\bibinfo {year} {2021})}\BibitemShut {NoStop}%
\bibitem [{\citenamefont {de~Neeve}\ \emph {et~al.}(2022)\citenamefont
  {de~Neeve}, \citenamefont {Nguyen}, \citenamefont {Behrle},\ and\
  \citenamefont {Home}}]{deNeeve2022Mar}%
  \BibitemOpen
  \bibfield  {author} {\bibinfo {author} {\bibfnamefont {Brennan}\ \bibnamefont
  {de~Neeve}}, \bibinfo {author} {\bibfnamefont {Thanh-Long}\ \bibnamefont
  {Nguyen}}, \bibinfo {author} {\bibfnamefont {Tanja}\ \bibnamefont {Behrle}},
  \ and\ \bibinfo {author} {\bibfnamefont {Jonathan~P.}\ \bibnamefont {Home}},\
  }\bibfield  {title} {\enquote {\bibinfo {title} {{Error correction of a
  logical grid state qubit by dissipative pumping}},}\ }\href {\doibase
  10.1038/s41567-021-01487-7} {\bibfield  {journal} {\bibinfo  {journal} {Nat.
  Phys.}\ }\textbf {\bibinfo {volume} {18}},\ \bibinfo {pages} {296--300}
  (\bibinfo {year} {2022})}\BibitemShut {NoStop}%
\bibitem [{\citenamefont {Ahn}\ \emph {et~al.}(2002)\citenamefont {Ahn},
  \citenamefont {Doherty},\ and\ \citenamefont {Landahl}}]{Ahn2002Mar}%
  \BibitemOpen
  \bibfield  {author} {\bibinfo {author} {\bibfnamefont {Charlene}\
  \bibnamefont {Ahn}}, \bibinfo {author} {\bibfnamefont {Andrew~C.}\
  \bibnamefont {Doherty}}, \ and\ \bibinfo {author} {\bibfnamefont {Andrew~J.}\
  \bibnamefont {Landahl}},\ }\bibfield  {title} {\enquote {\bibinfo {title}
  {{Continuous quantum error correction via quantum feedback control}},}\
  }\href {\doibase 10.1103/PhysRevA.65.042301} {\bibfield  {journal} {\bibinfo
  {journal} {Phys. Rev. A}\ }\textbf {\bibinfo {volume} {65}},\ \bibinfo
  {pages} {042301} (\bibinfo {year} {2002})}\BibitemShut {NoStop}%
\bibitem [{\citenamefont {Ahn}\ \emph {et~al.}(2003)\citenamefont {Ahn},
  \citenamefont {Wiseman},\ and\ \citenamefont {Milburn}}]{Ahn2003May}%
  \BibitemOpen
  \bibfield  {author} {\bibinfo {author} {\bibfnamefont {Charlene}\
  \bibnamefont {Ahn}}, \bibinfo {author} {\bibfnamefont {H.~M.}\ \bibnamefont
  {Wiseman}}, \ and\ \bibinfo {author} {\bibfnamefont {G.~J.}\ \bibnamefont
  {Milburn}},\ }\bibfield  {title} {\enquote {\bibinfo {title} {{Quantum error
  correction for continuously detected errors}},}\ }\href {\doibase
  10.1103/PhysRevA.67.052310} {\bibfield  {journal} {\bibinfo  {journal} {Phys.
  Rev. A}\ }\textbf {\bibinfo {volume} {67}},\ \bibinfo {pages} {052310}
  (\bibinfo {year} {2003})}\BibitemShut {NoStop}%
\bibitem [{\citenamefont {Sarovar}\ \emph {et~al.}(2004)\citenamefont
  {Sarovar}, \citenamefont {Ahn}, \citenamefont {Jacobs},\ and\ \citenamefont
  {Milburn}}]{Sarovar2004May}%
  \BibitemOpen
  \bibfield  {author} {\bibinfo {author} {\bibfnamefont {Mohan}\ \bibnamefont
  {Sarovar}}, \bibinfo {author} {\bibfnamefont {Charlene}\ \bibnamefont {Ahn}},
  \bibinfo {author} {\bibfnamefont {Kurt}\ \bibnamefont {Jacobs}}, \ and\
  \bibinfo {author} {\bibfnamefont {Gerard~J.}\ \bibnamefont {Milburn}},\
  }\bibfield  {title} {\enquote {\bibinfo {title} {{Practical scheme for error
  control using feedback}},}\ }\href {\doibase 10.1103/PhysRevA.69.052324}
  {\bibfield  {journal} {\bibinfo  {journal} {Phys. Rev. A}\ }\textbf {\bibinfo
  {volume} {69}},\ \bibinfo {pages} {052324} (\bibinfo {year}
  {2004})}\BibitemShut {NoStop}%
\bibitem [{\citenamefont {Sarovar}\ and\ \citenamefont
  {Milburn}(2005)}]{Sarovar2005Jul}%
  \BibitemOpen
  \bibfield  {author} {\bibinfo {author} {\bibfnamefont {Mohan}\ \bibnamefont
  {Sarovar}}\ and\ \bibinfo {author} {\bibfnamefont {G.~J.}\ \bibnamefont
  {Milburn}},\ }\bibfield  {title} {\enquote {\bibinfo {title} {{Continuous
  quantum error correction by cooling}},}\ }\href {\doibase
  10.1103/PhysRevA.72.012306} {\bibfield  {journal} {\bibinfo  {journal} {Phys.
  Rev. A}\ }\textbf {\bibinfo {volume} {72}},\ \bibinfo {pages} {012306}
  (\bibinfo {year} {2005})}\BibitemShut {NoStop}%
\bibitem [{\citenamefont {Wiseman}\ and\ \citenamefont
  {Milburn}(2014)}]{WisemanMilburnBook}%
  \BibitemOpen
  \bibfield  {author} {\bibinfo {author} {\bibfnamefont {Howard~M.}\
  \bibnamefont {Wiseman}}\ and\ \bibinfo {author} {\bibfnamefont {Gerard~J.}\
  \bibnamefont {Milburn}},\ }\href
  {https://www.amazon.com/Quantum-Measurement-Control-Howard-Wiseman/dp/1107424151}
  {\emph {\bibinfo {title} {{Quantum Measurement and Control}}}}\ (\bibinfo
  {publisher} {Cambridge University Press},\ \bibinfo {address} {Cambridge,
  England, UK},\ \bibinfo {year} {2014})\BibitemShut {NoStop}%
\bibitem [{\citenamefont {Mabuchi}(2009)}]{Mabuchi2009Oct}%
  \BibitemOpen
  \bibfield  {author} {\bibinfo {author} {\bibfnamefont {Hideo}\ \bibnamefont
  {Mabuchi}},\ }\bibfield  {title} {\enquote {\bibinfo {title} {{Continuous
  quantum error correction as classical hybrid control}},}\ }\href {\doibase
  10.1088/1367-2630/11/10/105044} {\bibfield  {journal} {\bibinfo  {journal}
  {New J. Phys.}\ }\textbf {\bibinfo {volume} {11}},\ \bibinfo {pages} {105044}
  (\bibinfo {year} {2009})}\BibitemShut {NoStop}%
\bibitem [{\citenamefont {Cardona}\ \emph {et~al.}(2019)\citenamefont
  {Cardona}, \citenamefont {Sarlette},\ and\ \citenamefont
  {Rouchon}}]{Cardona2019Jan}%
  \BibitemOpen
  \bibfield  {author} {\bibinfo {author} {\bibfnamefont {Gerardo}\ \bibnamefont
  {Cardona}}, \bibinfo {author} {\bibfnamefont {Alain}\ \bibnamefont
  {Sarlette}}, \ and\ \bibinfo {author} {\bibfnamefont {Pierre}\ \bibnamefont
  {Rouchon}},\ }\bibfield  {title} {\enquote {\bibinfo {title}
  {{Continuous-time Quantum Error Correction with Noise-assisted Quantum
  Feedback}},}\ }\href {\doibase 10.1016/j.ifacol.2019.11.778} {\bibfield
  {journal} {\bibinfo  {journal} {IFAC-PapersOnLine}\ }\textbf {\bibinfo
  {volume} {52}},\ \bibinfo {pages} {198--203} (\bibinfo {year}
  {2019})}\BibitemShut {NoStop}%
\bibitem [{\citenamefont {Mohseninia}\ \emph {et~al.}(2020)\citenamefont
  {Mohseninia}, \citenamefont {Yang}, \citenamefont {Siddiqi}, \citenamefont
  {Jordan},\ and\ \citenamefont {Dressel}}]{Mohseninia2020Nov}%
  \BibitemOpen
  \bibfield  {author} {\bibinfo {author} {\bibfnamefont {Razieh}\ \bibnamefont
  {Mohseninia}}, \bibinfo {author} {\bibfnamefont {Jing}\ \bibnamefont {Yang}},
  \bibinfo {author} {\bibfnamefont {Irfan}\ \bibnamefont {Siddiqi}}, \bibinfo
  {author} {\bibfnamefont {Andrew~N.}\ \bibnamefont {Jordan}}, \ and\ \bibinfo
  {author} {\bibfnamefont {Justin}\ \bibnamefont {Dressel}},\ }\bibfield
  {title} {\enquote {\bibinfo {title} {{Always-On Quantum Error Tracking with
  Continuous Parity Measurements}},}\ }\href {\doibase
  10.22331/q-2020-11-04-358} {\bibfield  {journal} {\bibinfo  {journal}
  {Quantum}\ }\textbf {\bibinfo {volume} {4}},\ \bibinfo {pages} {358}
  (\bibinfo {year} {2020})},\ \Eprint {http://arxiv.org/abs/1907.08882v2}
  {1907.08882v2} \BibitemShut {NoStop}%
\bibitem [{\citenamefont {Atalaya}\ \emph {et~al.}(2021)\citenamefont
  {Atalaya}, \citenamefont {Zhang}, \citenamefont {Niu}, \citenamefont
  {Babakhani}, \citenamefont {Chan}, \citenamefont {Epstein},\ and\
  \citenamefont {Whaley}}]{Atalaya2021Apr}%
  \BibitemOpen
  \bibfield  {author} {\bibinfo {author} {\bibfnamefont {J.}~\bibnamefont
  {Atalaya}}, \bibinfo {author} {\bibfnamefont {S.}~\bibnamefont {Zhang}},
  \bibinfo {author} {\bibfnamefont {M.~Y.}\ \bibnamefont {Niu}}, \bibinfo
  {author} {\bibfnamefont {A.}~\bibnamefont {Babakhani}}, \bibinfo {author}
  {\bibfnamefont {H.~C.~H.}\ \bibnamefont {Chan}}, \bibinfo {author}
  {\bibfnamefont {J.~M.}\ \bibnamefont {Epstein}}, \ and\ \bibinfo {author}
  {\bibfnamefont {K.~B.}\ \bibnamefont {Whaley}},\ }\bibfield  {title}
  {\enquote {\bibinfo {title} {{Continuous quantum error correction for
  evolution under time-dependent Hamiltonians}},}\ }\href {\doibase
  10.1103/PhysRevA.103.042406} {\bibfield  {journal} {\bibinfo  {journal}
  {Phys. Rev. A}\ }\textbf {\bibinfo {volume} {103}},\ \bibinfo {pages}
  {042406} (\bibinfo {year} {2021})}\BibitemShut {NoStop}%
\bibitem [{\citenamefont {Livingston}\ \emph {et~al.}(2022)\citenamefont
  {Livingston}, \citenamefont {Blok}, \citenamefont {Flurin}, \citenamefont
  {Dressel}, \citenamefont {Jordan},\ and\ \citenamefont
  {Siddiqi}}]{Livingston2021Jul}%
  \BibitemOpen
  \bibfield  {author} {\bibinfo {author} {\bibfnamefont {William~P.}\
  \bibnamefont {Livingston}}, \bibinfo {author} {\bibfnamefont {Machiel~S.}\
  \bibnamefont {Blok}}, \bibinfo {author} {\bibfnamefont {Emmanuel}\
  \bibnamefont {Flurin}}, \bibinfo {author} {\bibfnamefont {Justin}\
  \bibnamefont {Dressel}}, \bibinfo {author} {\bibfnamefont {Andrew~N.}\
  \bibnamefont {Jordan}}, \ and\ \bibinfo {author} {\bibfnamefont {Irfan}\
  \bibnamefont {Siddiqi}},\ }\bibfield  {title} {\enquote {\bibinfo {title}
  {{Experimental demonstration of continuous quantum error correction}},}\
  }\href {\doibase 10.1038/s41467-022-29906-0} {\bibfield  {journal} {\bibinfo
  {journal} {Nat. Commun.}\ }\textbf {\bibinfo {volume} {13}},\ \bibinfo
  {pages} {1--7} (\bibinfo {year} {2022})}\BibitemShut {NoStop}%
\bibitem [{\citenamefont {Di{\ifmmode\acute{o}\else\'{o}\fi}si}\ \emph
  {et~al.}(2006)\citenamefont {Di{\ifmmode\acute{o}\else\'{o}\fi}si},
  \citenamefont {Konrad}, \citenamefont {Scherer},\ and\ \citenamefont
  {Audretsch}}]{Diosi2006Sep}%
  \BibitemOpen
  \bibfield  {author} {\bibinfo {author} {\bibfnamefont {Lajos}\ \bibnamefont
  {Di{\ifmmode\acute{o}\else\'{o}\fi}si}}, \bibinfo {author} {\bibfnamefont
  {Thomas}\ \bibnamefont {Konrad}}, \bibinfo {author} {\bibfnamefont {Artur}\
  \bibnamefont {Scherer}}, \ and\ \bibinfo {author} {\bibfnamefont
  {J{\ifmmode\ddot{u}\else\"{u}\fi}rgen}\ \bibnamefont {Audretsch}},\
  }\bibfield  {title} {\enquote {\bibinfo {title} {{Coupled Ito equations of
  continuous quantum state measurement and estimation}},}\ }\href {\doibase
  10.1088/0305-4470/39/40/l01} {\bibfield  {journal} {\bibinfo  {journal} {J.
  Phys. A: Math. Gen.}\ }\textbf {\bibinfo {volume} {39}},\ \bibinfo {pages}
  {L575--L581} (\bibinfo {year} {2006})}\BibitemShut {NoStop}%
\bibitem [{\citenamefont {Borah}\ \emph {et~al.}(2021)\citenamefont {Borah},
  \citenamefont {Sarma}, \citenamefont {Kewming}, \citenamefont {Milburn},\
  and\ \citenamefont {Twamley}}]{Borah2021Nov}%
  \BibitemOpen
  \bibfield  {author} {\bibinfo {author} {\bibfnamefont {Sangkha}\ \bibnamefont
  {Borah}}, \bibinfo {author} {\bibfnamefont {Bijita}\ \bibnamefont {Sarma}},
  \bibinfo {author} {\bibfnamefont {Michael}\ \bibnamefont {Kewming}}, \bibinfo
  {author} {\bibfnamefont {Gerard~J.}\ \bibnamefont {Milburn}}, \ and\ \bibinfo
  {author} {\bibfnamefont {Jason}\ \bibnamefont {Twamley}},\ }\bibfield
  {title} {\enquote {\bibinfo {title} {{Measurement-Based Feedback Quantum
  Control with Deep Reinforcement Learning for a Double-Well Nonlinear
  Potential}},}\ }\href {\doibase 10.1103/PhysRevLett.127.190403} {\bibfield
  {journal} {\bibinfo  {journal} {Phys. Rev. Lett.}\ }\textbf {\bibinfo
  {volume} {127}},\ \bibinfo {pages} {190403} (\bibinfo {year}
  {2021})}\BibitemShut {NoStop}%
\bibitem [{\citenamefont {Minev}\ \emph {et~al.}(2019)\citenamefont {Minev},
  \citenamefont {Mundhada}, \citenamefont {Shankar}, \citenamefont {Reinhold},
  \citenamefont
  {Guti{\ifmmode\acute{e}\else\'{e}\fi}rrez-J{\ifmmode\acute{a}\else\'{a}\fi}uregui},
  \citenamefont {Schoelkopf}, \citenamefont {Mirrahimi}, \citenamefont
  {Carmichael},\ and\ \citenamefont {Devoret}}]{Minev2019Jun}%
  \BibitemOpen
  \bibfield  {author} {\bibinfo {author} {\bibfnamefont {Z.~K.}\ \bibnamefont
  {Minev}}, \bibinfo {author} {\bibfnamefont {S.~O.}\ \bibnamefont {Mundhada}},
  \bibinfo {author} {\bibfnamefont {S.}~\bibnamefont {Shankar}}, \bibinfo
  {author} {\bibfnamefont {P.}~\bibnamefont {Reinhold}}, \bibinfo {author}
  {\bibfnamefont {R.}~\bibnamefont
  {Guti{\ifmmode\acute{e}\else\'{e}\fi}rrez-J{\ifmmode\acute{a}\else\'{a}\fi}uregui}},
  \bibinfo {author} {\bibfnamefont {R.~J.}\ \bibnamefont {Schoelkopf}},
  \bibinfo {author} {\bibfnamefont {M.}~\bibnamefont {Mirrahimi}}, \bibinfo
  {author} {\bibfnamefont {H.~J.}\ \bibnamefont {Carmichael}}, \ and\ \bibinfo
  {author} {\bibfnamefont {M.~H.}\ \bibnamefont {Devoret}},\ }\bibfield
  {title} {\enquote {\bibinfo {title} {{To catch and reverse a quantum jump
  mid-flight - Nature}},}\ }\href {\doibase 10.1038/s41586-019-1287-z}
  {\bibfield  {journal} {\bibinfo  {journal} {Nature}\ }\textbf {\bibinfo
  {volume} {570}},\ \bibinfo {pages} {200--204} (\bibinfo {year}
  {2019})}\BibitemShut {NoStop}%
\bibitem [{\citenamefont {Blais}\ \emph {et~al.}(2021)\citenamefont {Blais},
  \citenamefont {Grimsmo}, \citenamefont {Girvin},\ and\ \citenamefont
  {Wallraff}}]{Blais2021May}%
  \BibitemOpen
  \bibfield  {author} {\bibinfo {author} {\bibfnamefont {Alexandre}\
  \bibnamefont {Blais}}, \bibinfo {author} {\bibfnamefont {Arne~L.}\
  \bibnamefont {Grimsmo}}, \bibinfo {author} {\bibfnamefont {S.~M.}\
  \bibnamefont {Girvin}}, \ and\ \bibinfo {author} {\bibfnamefont {Andreas}\
  \bibnamefont {Wallraff}},\ }\bibfield  {title} {\enquote {\bibinfo {title}
  {{Circuit quantum electrodynamics}},}\ }\href {\doibase
  10.1103/RevModPhys.93.025005} {\bibfield  {journal} {\bibinfo  {journal}
  {Rev. Mod. Phys.}\ }\textbf {\bibinfo {volume} {93}},\ \bibinfo {pages}
  {025005} (\bibinfo {year} {2021})}\BibitemShut {NoStop}%
\end{thebibliography}

% \newpage
% \input{supplementary}
%TC:endignore

%merlin.mbs apsrev4-1.bst 2010-07-25 4.21a (PWD, AO, DPC) hacked
%Control: key (0)
%Control: author (0) dotless jnrlst
%Control: editor formatted (1) identically to author
%Control: production of article title (0) allowed
%Control: page (1) range
%Control: year (0) verbatim
%Control: production of eprint (0) enabled
%

\end{document}